\documentclass{article}
\usepackage[utf8]{inputenc}

\usepackage[affil-it]{authblk}
\usepackage{graphicx}
\usepackage{amsmath}
\usepackage{amssymb}
\usepackage{amsthm}
\usepackage{bm}
\usepackage{geometry}
\usepackage{hyperref}
\hypersetup{
    colorlinks = true,
    linkcolor = blue,
    citecolor = blue,
    urlcolor  = blue
}
\usepackage{mathtools}
\usepackage{cases}
\usepackage{color}
\usepackage{makecell}
\usepackage{multicol}
\usepackage{multirow}
\usepackage{tabularx}
\usepackage{subcaption}
\usepackage{chemformula}
\usepackage[titletoc,page]{appendix}

\usepackage{titlesec}
\usepackage{esint}
\titleformat{\subsection}{\large\bfseries}{\thesubsection}{1em}{}
\titleformat{\subsubsection}{\large\bfseries}{\thesubsubsection}{1em}{}

\newcolumntype{C}{>{\centering\arraybackslash}X}

\newcommand{\dd}{\mathrm{d}}
\newcommand{\del}{\partial}
\newcommand{\ii}{\mathrm{i}}
\newcommand{\fdd}[2]{\frac{\dd #1}{\dd #2}}         % dx/dy
\newcommand{\fpp}[2]{\frac{\del #1}{\del #2}}       % (del x)/(del y)
\newcommand{\D}{\displaystyle}

\newcommand{\K}{\,\text{K}}                         % space + Kelvin
\newcommand{\m}{\,\text{m}}                         % space + m
\newcommand{\cm}{\,\text{cm}}                         % space + cm
\newcommand{\mm}{\,\text{mm}}                         % space + mm
\newcommand{\nm}{\,\text{nm}}                         % space + nm

\newcommand{\Ei}{\operatorname{Ei}}                 % exponential integral function
\newcommand{\zR}{z_{\text{R}}}                      % Rayleigh range
\newcommand{\eff}{_{\text{eff}}}                    % eff subscript

\title{\bfseries A study of enhancing the power of a CW Ti:sapphire laser by cryogenic cooling}

\author{Simai Jia}
\affil{Clarendon Laboratory, University of Oxford, Parks Road, Oxford, OX1 3PU, UK}
\date{December 2022}

\begin{document}

\maketitle

\large

\tableofcontents

% Sec1_introduction.tex
% cannot be complied alone

\section{Introduction}
\subsection{The need for high laser power}
% Existing schemes that use two photons cannot be scaled to the required size, e.g.\ Raman transitions  that impart momentum of $2\hbar k$ in a coherent transfer between states in the ground configuration of alkali atoms. To be suitable for matter-wave interferometry, the single-photon transition must have a long-lived upper level so that there is a very low probability of spontaneous emission during multiple-pulse sequences that coherently drive atoms up and down many times; this occurs for so-called clock transitions that have an extremely narrow linewidth. Consequently these clock transitions also have small electric-dipole matrix elements and require a high laser intensity to achieve a high Rabi frequency.

The Matter-wave Atomic Gradiometer Interferometric Sensor (MAGIS-100) and Atom Interferometer Observatory and Network (AION) projects will use single-photon transitions for atom interferometry to enable long baselines \cite{Abe:2021,Badurina:2020}, in a way that is not feasible with existing two-photon techniques. The single-photon approach requires using a clock transition from the ground state to a long-lived  excited level so that negligible spontaneous emission occurs while atoms reside in the upper level for a short but finite time between up and down momentum transfer pulses. This is a real population inversion whereas the intermediate level in a Raman transition, or Bragg diffraction, is virtual.

Clock transitions are very weak, ‘forbidden’, and therefore require a strong laser beam to give a high Rabi frequency, e.g.\ a few Watts with a cm-scale laser beam waist in the first instance, and powers of  100 Watts are desirable to realize the full advantage of this approach, as explained in this section. This beam waist is necessary to give flat wave fronts over a long distance in large-scale atom interferometers. Comparable power is used nowadays to obtain good performance in the atom interferometry experiments with Rb atoms at Stanford --- a large frequency detuning is used for the Raman transitions to reduce spontaneous emission during the process.

As well as wanting to have short pulses so that many thousands can be fitted into the limited time of flight (e.g.\ $T_{\text{TOF}} = 2$\,s and a splitting-deflection-recombination sequence of 4000 pulses (1:2:1) gives a maximum pulse length of 500\,$\mu$s), there is a less obvious reason for wanting a high Rabi frequency, namely to minimize the inefficiencies in the $\pi$-pulses arising from non-zero frequency detunings \cite{Abe:2021}. Frequency shifts of the laser radiation relative to the transition arise from laser frequency noise or Doppler broadening caused by a velocity spread in the atom cloud. The extreme narrowness of the clock transition means that even small frequency detunings affect the efficiency of $\pi$-pulses and require the laser frequency to be stabilized to “within 10 Hz of the clock transition resonance using the optical frequency comb” \cite{Abe:2021}. Shorter $\pi$-pulses mitigate the effect of laser frequency shifts by increasing the Fourier-transform limited bandwidth, and it is beneficial to have sufficient power to reach Rabi frequencies at this level in order to minimize $\pi$-pulse inefficiencies arising from nonzero detunings.

In the paper detailing the MAGIS proposal~\cite{Abe:2021}, it is assumed that large momentum transfer (LMT) atom interferometry based on the 698\,nm transition of strontium requires 8\,W of continuous wave (CW) power at this wavelength to give a Rabi frequency of several kHz on the clock transition with a cm-scale laser beam waist. In contrast, optical-lattice clocks use a long interrogation time to obtain a signal that is very sensitive to the frequency difference between the radiation and the transition. Since the Rabi frequency is proportional to the electric field strength of the electromagnetic radiation which scales as the square root of the intensity, in round numbers, we assume 100\,W as a goal for the laser power that we want to achieve for MAGIS-100 or AION in the future.

\subsection[How to achieve 100 W of red laser radiation?]{How to achieve 100\,W of red laser radiation?}
As outlined in the previous section, atom interferometry on the 698\,nm clock can be implemented using around 8\,W of radiation with a bandwidth of 10\,Hz but increasing this to 100\,W is a goal for the future. 

A power of 8\,W at 698\,nm can be produced by a system of two titanium-doped sapphire (Ti:sapphire or Ti:Sa) systems operated as an amplifier which is injection-locked by a narrow-bandwidth laser (as reported by M Squared Lasers at the first AION workshop, see Ref.~\cite{Thom:2019}). In this pairing both the laser and amplifier are pumped by separate 18\,W green lasers, but a narrow bandwidth laser can be used to injection-lock more than one amplifier, e.g.\ 12 amplifiers of 8\,W each giving a total of 96\,W could be seeded by 2 or 3 lasers, making a total of around 15 systems and pump lasers. 

\begin{figure}[htbp]
    \hspace{.15\textwidth}
    \includegraphics[width = .6\textwidth]{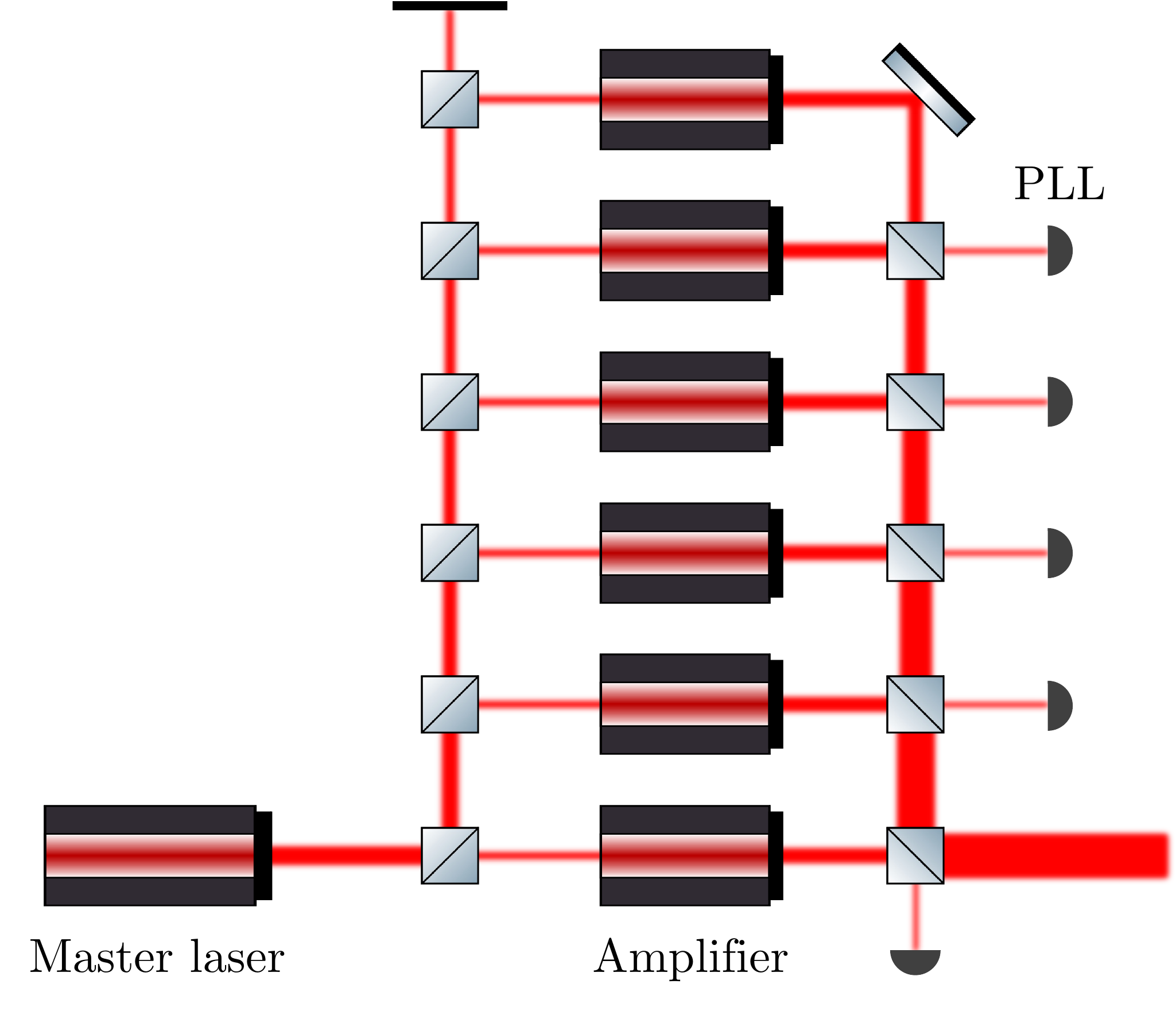}
    \hfill
    \caption{Simplified schematic to produce high power laser by coherent addition. PLL : phase-locked loop. The master laser and amplifiers each have a pump laser and an optical cavity with Ti:Sa gain medium. The injection-locked amplifiers do not need the intracavity elements (etalons etc.) necessary for a single-frequency laser oscillator, and therefore have lower round-trip loss and higher output power (by a factor of 2 typically). Such a system might be expected to give a power of 48\,W, extrapolating from currently achieved results.}
    \label{fig:coherent_combination}
\end{figure}

In principle, the output of these systems can be combined by coherent addition on a beam splitter as illustrated in Figure~\ref{fig:coherent_combination}; the coherent addition of laser light is used to generate light for the Rb Raman experiments at Stanford (from a few lasers)~\cite{Giunta:2021}. Constructive interference at the beam splitter requires phase-locking of the two incoming beams which can be achieved by sophisticated but standard electronics, and this ensures that the output beam maintains the same narrow bandwidth of the incident light. Scaling to high power using 15, or more, commercial units (Ti:Sa plus pump laser) has the advantage of being a ‘straightforward’ extension of known techniques, not requiring extensive development, but it is worthwhile to consider alternatives (especially given the reliability issues that have already been encountered by both MAGIS-100 and AION, with pump lasers).

\subsection{Cryogenic cooling of titanium-doped sapphire}
Titanium-doped sapphire has several properties that make it a very good gain medium for laser amplification. It provides gain over a broad spectral region, making it very suitable for ultrafast laser systems (femtosecond pulses in the time domain correspond to a wide range of frequencies); although we use only a small part of that range, in the red, this means that very high-quality crystals are available as are used in extremely high-power laser facilities where the intensities far exceed those in CW systems, so the damage threshold is irrelevant. Rather the limitation to the power in both pulsed and CW applications comes from thermal effects --- removing the heat dumped into the gain medium by the pump laser beam. 

\begin{figure}[htbp]
    \centering
    \includegraphics[width = .6\textwidth]{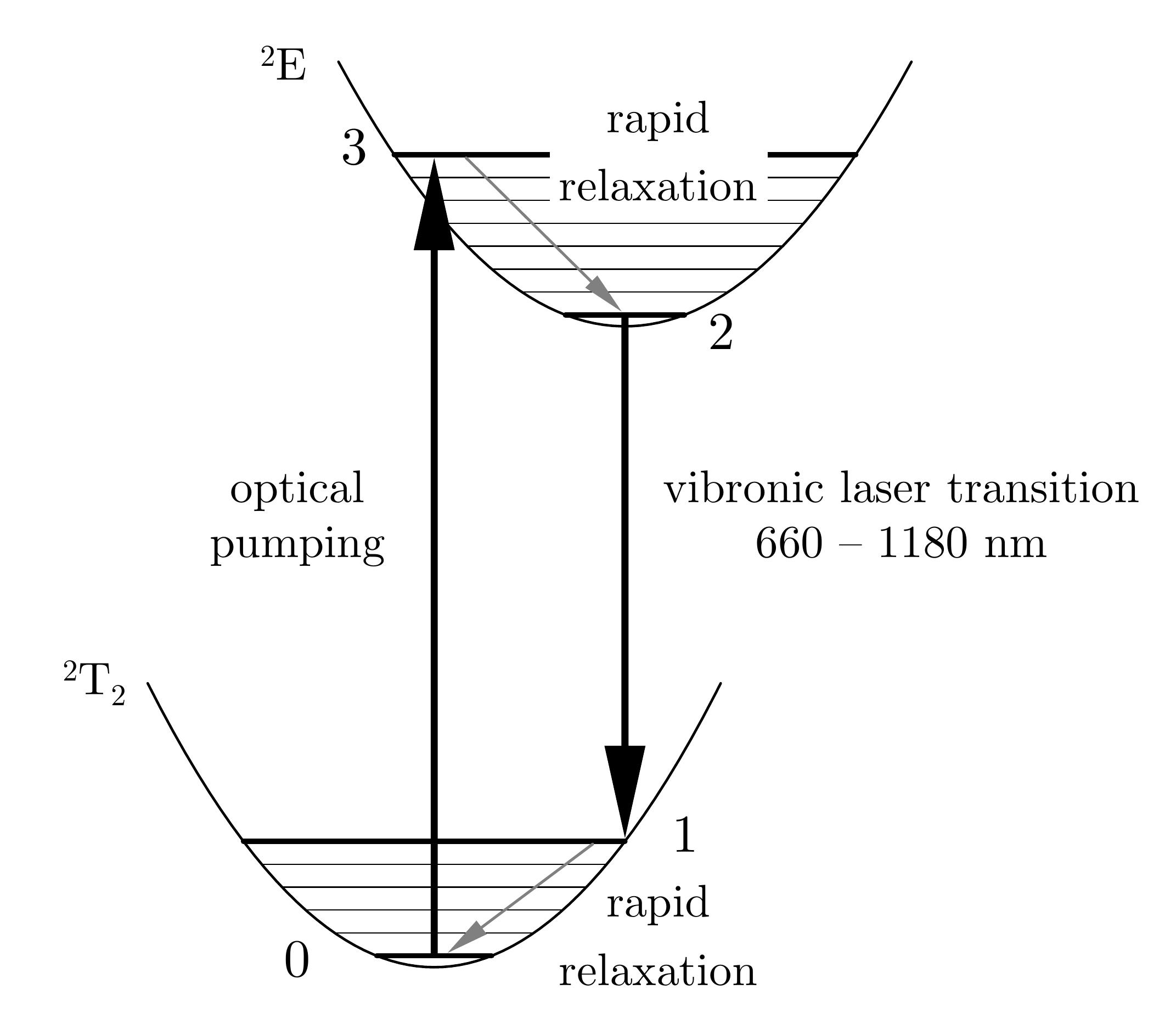}
    \caption{Simplified energy level structure and laser transitions of the \ch{Ti^3+} ion in \ch{Al2O3} (Ti:sapphire).}
    \label{fig:ti:sa_transition}
\end{figure}

Ti:Sa is a four-level laser scheme as illustrated in Figure~\ref{fig:ti:sa_transition} and for each stimulated photon emitted at 698\,nm the difference in energy between it and the pump (at 532\,nm) goes into vibrational energy of the crystal lattice (an intrinsic inefficiency of 24\%, or a ‘quantum efficiency’ of 76\%). In practice, the pump beam is not completely absorbed on passing through the laser crystal and will not perfectly overlap with the laser mode but the observed efficiency of 44\% is high (8\,W for 18\,W input --- this is the minimum specification of the pump laser and the output may be about 20\,W initially). Typically, the output power of a laser is linearly proportional to the pump power above the lasing threshold but this linear growth rolls off when thermal lensing starts to affect the optical cavity. 
% [Check whether we have any data on this for our laser system?] 

This project explores how cryogenic cooling of the laser crystal greatly reduces thermal lensing (by over an order of magnitude) because of two factors: (i) the increase in the thermal conductivity of the crystal (as the phonon density decreases --- the impurities from the titanium doping are not significant), and (ii) the decrease in the temperature dependence of the refractive index. The advantage of cooling is used in state-of-the-art pulsed laser systems (further details in Appendix~\ref{app:dipole_project}). The gain curve also changes with temperature, becoming slightly less broad, but this is not a significant effect at 698\,nm~\cite{Burton:2017}. Thus, it might be possible to obtain 100\,W at 698\,nm from a single amplifier with a suitable pump laser. The aim of this report is to investigate the physics that may be a factor in choosing between having $12\times 8$\,W, $3\times 30$\,W, or $1\times 100$\,W etc. Other factors will be commercial availability of components, cost and reliability (see Appendices~\ref{app:companies_and_project}, \ref{app:pump_lasers} and \ref{app:cost_benefit} for further discussion).

\clearpage
% Sec2_temperature_profile.tex
% cannot be complied alone

\section{Transverse Profile of Temperature}

\subsection{The problem}
Heated by the pump laser at the center and cooled at the surroundings, there will be a temperature distribution in the Ti:sapphire crystal. At thermal equilibrium, the heat dumped by the pump laser is dissipated by the heat flow caused by the temperature gradient inside the crystal. The temperature field follows the equation
\begin{equation}
    - \kappa \nabla^2 T = q(\mathbf{r})
    \label{eq:del_temp}
\end{equation}
where $\kappa$ is the thermal conductivity (assumed to be a constant inside the crystal) and $q(\mathbf{r})$ is the heating power per unit volume at position $\mathbf{r}$.

We assume that the boundary of the crystal is at the same temperature as the environment, either at room temperature (RT) or at some cryogenic temperature. As such, the boundary associated to Eq.~(\ref{eq:del_temp}) is that
\begin{equation}
    T = T_0 \qquad \text{at } r = R
\end{equation}
where $T_0$ is the environment temperature and $R$ is the radius of the laser crystal.

In the following subsections, we neglect boundary effects at the front and rear polished surfaces, assume that the pump is uniform along the laser crystal and solve for the temperature distribution. Under such circumstances, the problem has translational symmetry along the central axis. We take into account two scenarios: (i) the pump beam inside the crystal is circular; (ii) the pump beam inside the crystal is elliptical. While analytical solution is obtainable for the first case, the latter can only be evaluated numerically.

\subsection{Circular Gaussian beam source}
With the pump beam inside the laser crystal being a circular Gaussian beam, we also have rotational symmetry about the central axis.

The heating power per unit volume $q(\mathbf{r})$ is a certain fraction of the energy density of the pump beam of which the transverse profile is
\begin{equation*}
    \D I(r) = I_0 \exp\left( -\frac{2 r^2}{w_0^2} \right)
\end{equation*}
where $w_0$ is the $1/e^2$-beam radius. Therefore, we can write
\begin{equation}
    \D q(r) = q_0 \exp\left( -\frac{r^2}{w_1^2} \right).
    \label{eq:source_gaus}
\end{equation}
where $w_1^2 = w_0^2/2$.

We solve for the temperature field under cylindrical coordinates of which the $z-$axis coincide with the central axis of the crystal. Because of the translational symmetry along $z$ and rotational symmetry about $z$, the temperature should be a function of the radial coordinate $r$ only. So we have that
\begin{equation}
    - \kappa\, \frac{1}{r}\fdd{}{r}\left(r\,\fdd{T}{r} \right) = q(r)\ .
\end{equation}
Integrating this equation once, we get
\begin{equation}
    r\, \fdd{T}{r} = -\frac{Q_0}{2\pi\kappa} \left[ 1- \exp\left( -\frac{r^2}{w_1^2} \right) \right]
    \label{eq:temp_r}
\end{equation}
where $Q_0 = \pi q_0 w_1^2$ is the total heat dumped by the pump laser per unit time per unit length. Solving for $T(r)$ with the boundary condition $T(r=R) = T_0$, the solution can be written as
\begin{equation}
    T(r) = T_0 + T_p\, g(r)
\end{equation}
where
\begin{equation}
    T_p  = \frac{Q_0}{2\pi\kappa} = \frac{q_0 w_1^2}{2 \kappa}
    \label{eq:Tp_def}
\end{equation}
is a quantity with dimensions of temperature that characterizes heating by the pump laser and
\begin{equation}
    g(r) = \ln\frac{R}{r} + \frac12\left[ \Ei\left( -\frac{r^2}{w_1^2} \right)
    -\Ei\left( -\frac{R^2}{w_1^2} \right) \right]
\end{equation}
is a dimensionless function that describes the radial distribution of excess temperature, with $\Ei(x)$ being the exponential integral defined as
\begin{equation*}
    \Ei(x) = \int_{-\infty}^x \frac{e^t}{t}\,\dd t\ .
\end{equation*}

From this, we can calculate the temperature difference between the center and the boundary of the crystal which is given by
\begin{equation}
    \Delta T = T_p\, (g(0) - g(R)) = T_p \left( \ln\frac{R}{w_1} + \frac12\,\gamma \right)
\end{equation}
where $\gamma \approx 0.577$ is the Euler's constant. For a typical laser crystal with $R = 4\,\text{mm}$, $L = 20\,\text{mm}$ and $\kappa = 35\,\text{W}\text{m}^{-1}\text{K}^{-1}$ at RT, absorbing heat of $20\,\text{W}$ and assuming $R/w_1 = 10$, we have $\Delta T = 12\K$.

\begin{figure}[ht]
    \centering
    \includegraphics[width=.48\textwidth]{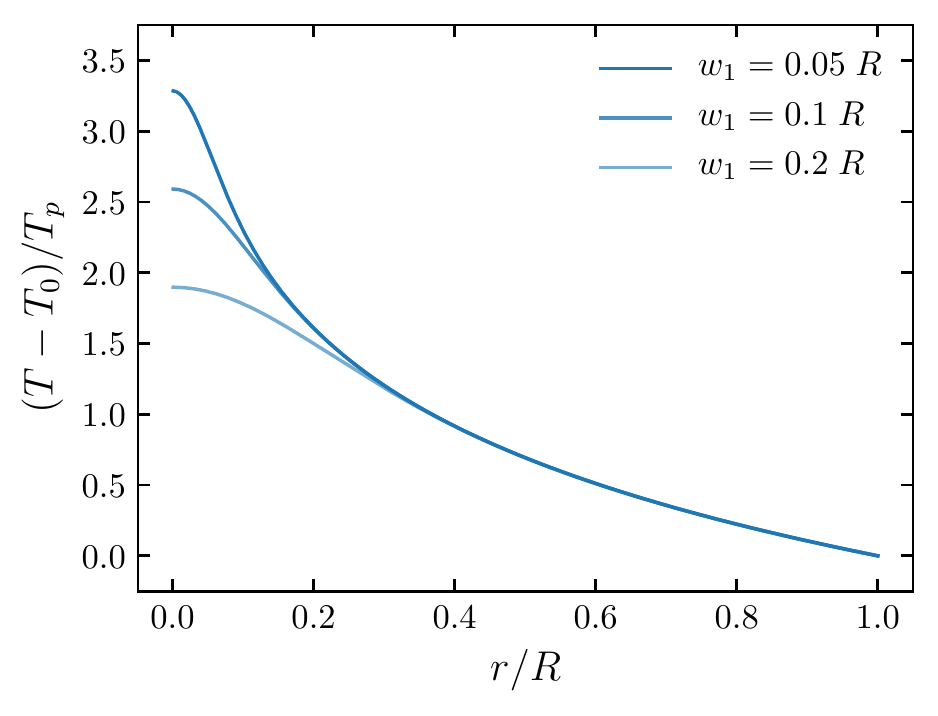}

    \caption{The dimensionless function $g(r)$ plotted for different $w_1$.}
    \label{fig:gaus_r_compare_w}
\end{figure}

To visualize the solution we get, in Figure~\ref{fig:gaus_r_compare_w}, the function $g(r)$ is plotted for different values of $w_1$. As $w_1$ increases (i.e.\ the pump beam expands while remaining at the same power), the temperature near the center of the crystal decreases but remains at the same level towards the boundary.

\begin{figure}[hb]
    \begin{subfigure}[b]{.48\textwidth}
    \centering
    \includegraphics[width=\textwidth]{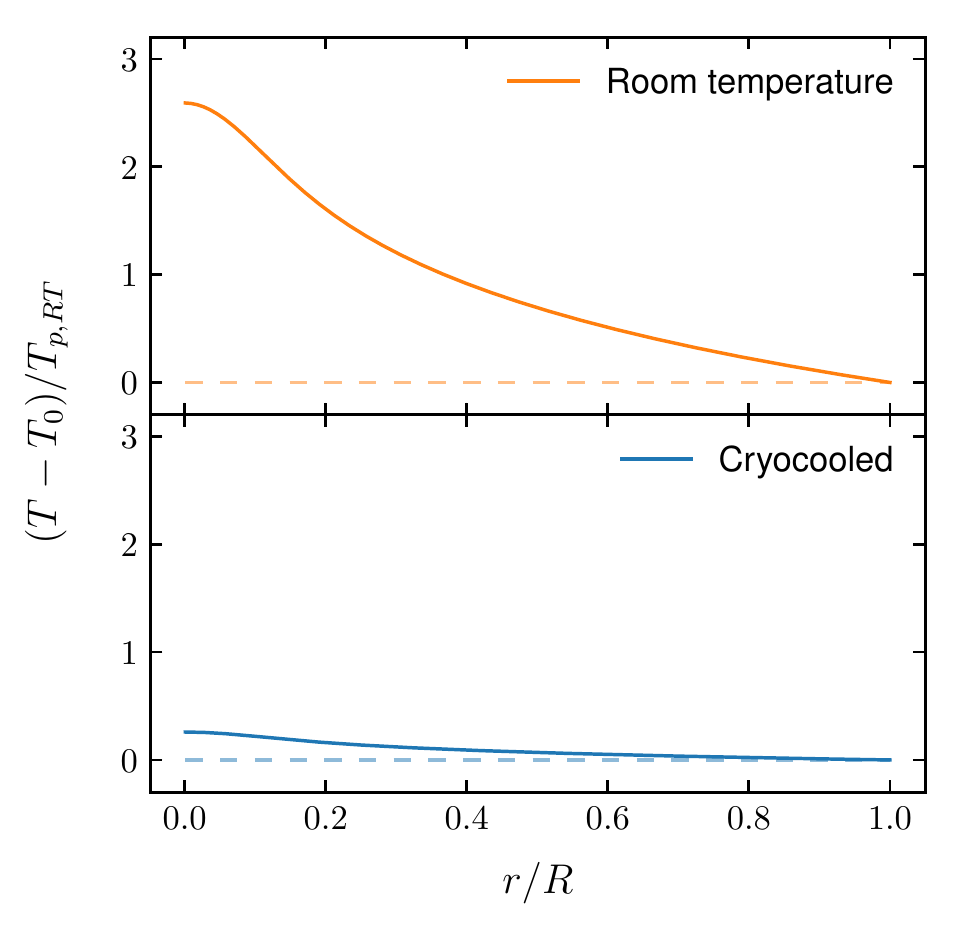}
    \subcaption{$w_1 = 0.1\ R$}
    \end{subfigure}
    \hfill
    \begin{subfigure}[b]{.48\textwidth}
    \centering
    \includegraphics[width=\textwidth]{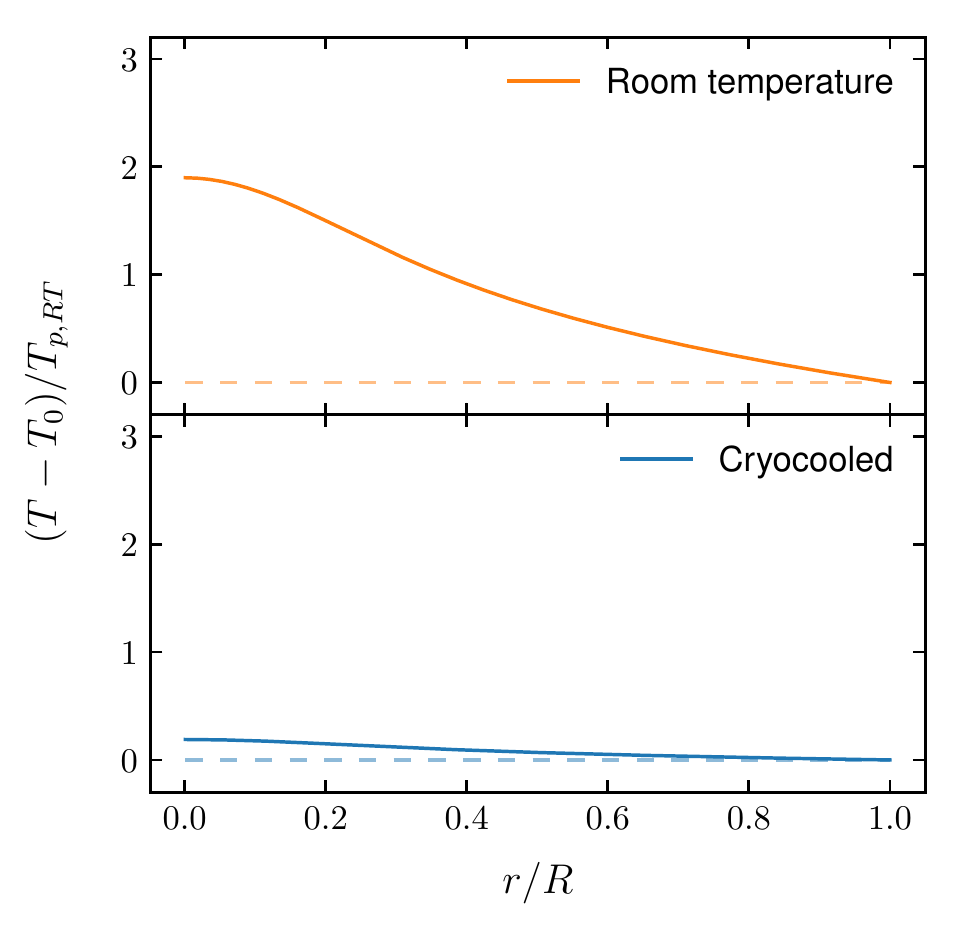}
    \subcaption{$w_1 = 0.2\ R$}
    \end{subfigure}
    
    \caption{Radial temperature profiles under different cooling conditions and different pump beam widths. $T_0 = 300\K$ for normal room temperature condition (upper) and $T_0 = 100\K$ for cryocooled scenario (lower).}
    \label{fig:temp_r_gaus}
\end{figure}

In Figure~\ref{fig:temp_r_gaus}, we compare the temperature profiles under different cooling conditions and different width of the pump beam. The crystal is supposed to be either in room temperature environment ($T_0 = 300\K$) or cooled cryogenically (here we take $T_0 = 100\K$). The thermal conductivity vs.\ temperature relation of sapphire crystal in Appendix~\ref{app:thermal_index} shows that
\begin{equation}
    \kappa(T = 100\K) \approx 10 \, \kappa(T = 300\K)\ .
    \notag
\end{equation}
Therefore
\begin{equation*}
    T_p(T_0 = 100\K) \approx \frac{1}{10}\,T_p(T_0 = 300\K)\ .
\end{equation*}
Cooling of the crystal to cryogenic temperatures will dramatically reduce the temperature unevenness and thus elongate the thermal lens, as can be seen by comparing the orange (RT) and blue (cryogenic) curves. On the other hand, if the pump beam is expanded while its total power is fixed, then the parameter $T_p$ will not change but the temperature unevenness is still mitigated. Quantitative results for the thermal lens are presented in the next section.

\clearpage

\clearpage
\subsection{Beam expansion through refraction}
In the previous subsection, we assumed that the pump beam is circular inside the laser crystal. If, however, the pump beam is circular before entering the laser crystal, then it will become elliptical because of refraction.

\begin{figure}[h!]
    \centering
    \includegraphics[width = .85\textwidth]{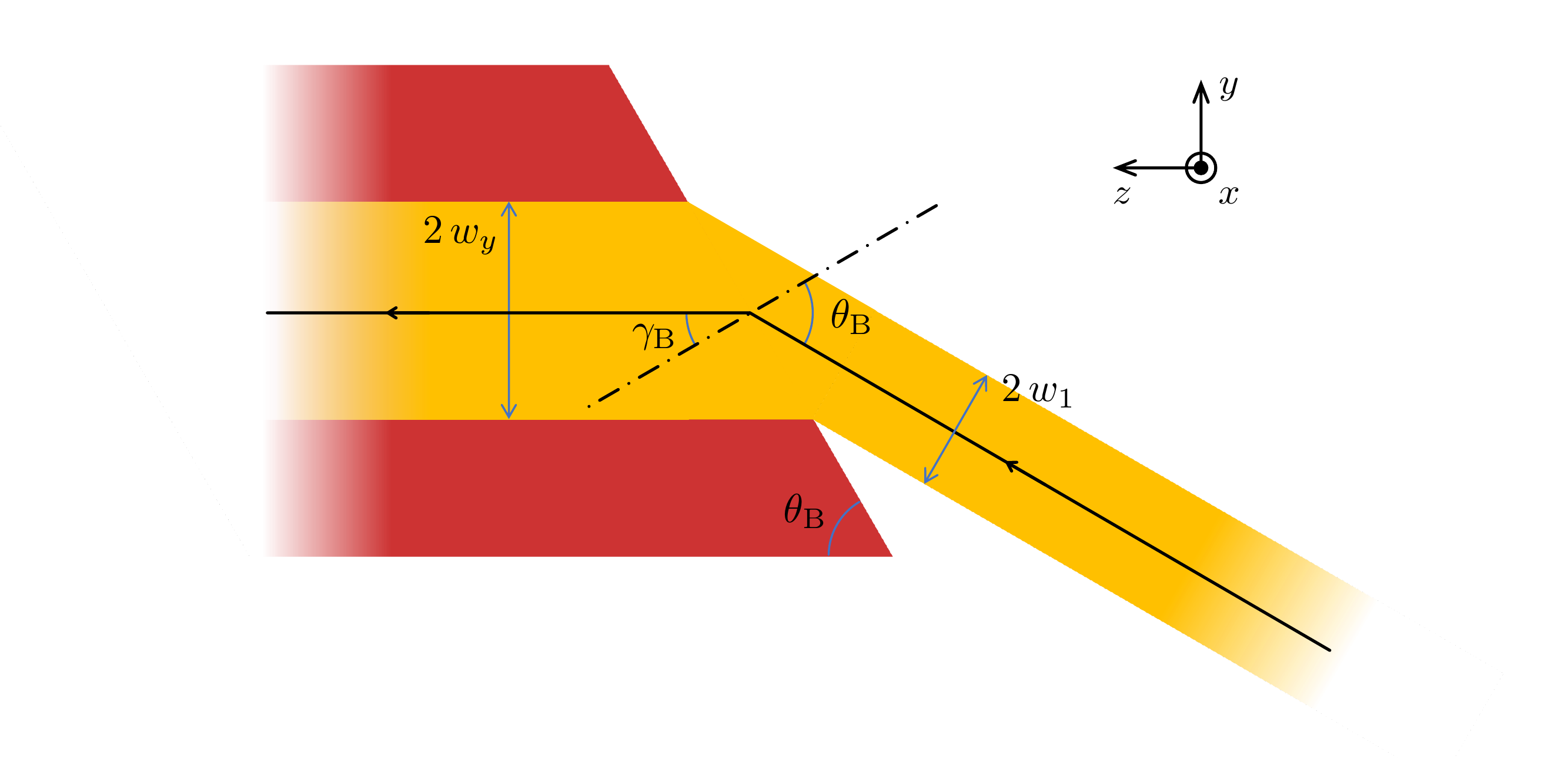}
    \caption{Refraction of the pump laser beam upon entering the laser crystal. The incident angle is Brewster's angle $\theta_{\text{B}} = \arctan n = 60.4^{\circ}$ and the refraction angle is $\gamma_{\text{B}} = \arctan (1/n) = 29.6^{\circ}$.}
    \label{fig:refraction}
\end{figure}

As shown in Figure~\ref{fig:refraction}, in the plane where the laser beam inside the cavity circulates (which we call the horizontal plane), the width of the pump beam is projected twice upon entering the laser crystal. In comparison, its width in the direction perpendicular to this plane remains unchanged. As the beam is incident on the crystal at Brewster's angle $\theta_{\text{B}} = \arctan n$, if we still record the beam width in air as $w_1$, then the beam widths in the $x$ (vertical) and $y$ (horizontal) directions are given by
\begin{align}
    w_{x} &= w_1 \\
    w_{y} &= w_1 \tan\theta_{\text{B}} = n w_1 \ .
\end{align}

The intensity profile of an elliptical Gaussian beam with widths $w_x$ and $w_y$ is given by
\begin{equation}
    I(x,y) = I_0 \exp\left[ -\left( \frac{x^2}{w_x^2} + \frac{y^2}{w_y^2} \right) \right]
    \notag
\end{equation}
and thus the heating power per unit volume
\begin{equation}
    q(x,y) = q_0 \exp\left[ -\left( \frac{x^2}{w_x^2} + \frac{y^2}{w_y^2} \right) \right]\ .
    \label{eq:source_ellip}
\end{equation}
This makes the laser crystal have different thermal lenses in the two directions. Therefore, the thermal lens formed by the laser crystal is not a circular one, but a lens with astigmatism.

\subsection{Elliptical Gaussian beam source}
With the source function given by Eq.~(\ref{eq:source_ellip}), the temperature now follows
\begin{equation}
    -\kappa\left(\fpp{^2}{x^2} + \fpp{^2}{y^2}\right) T = q_0 \exp\left[ -\left( \frac{x^2}{w_x^2} + \frac{y^2}{w_y^2} \right) \right]
    \label{eq:temp_xy}
\end{equation}
with the boundary condition given by
\begin{equation}
    T(x,y) = T_0 \qquad \text{at } x^2 + y^2 = R^2 \ .
    \label{eq:temp_xy_boundary}
\end{equation}
Making change of variables
\begin{equation}
    u = \frac{\kappa}{q_0}\,(T - T_0)\ ,
    \label{eq:u_to_T}
\end{equation}
where $u$ has the dimension of $[L]^2$, the problem becomes
\begin{numcases}
    \D -\left(\fpp{^2}{x^2} + \fpp{^2}{y^2}\right) u = \exp\left[ -\left( \frac{x^2}{w_x^2} + \frac{y^2}{w_y^2} \right) \right]
    \label{eq:u_xy} \\
    \notag \vspace{3mm} \\
    \, u(x,y) = 0 \qquad \text{at } x^2 + y^2 = R^2
    \label{eq:u_xy_boundary}
\end{numcases}
which can be solved numerically.

\begin{figure}[htb]
    \begin{subfigure}[b]{.48\textwidth}
    \centering
    \includegraphics[width=\textwidth]{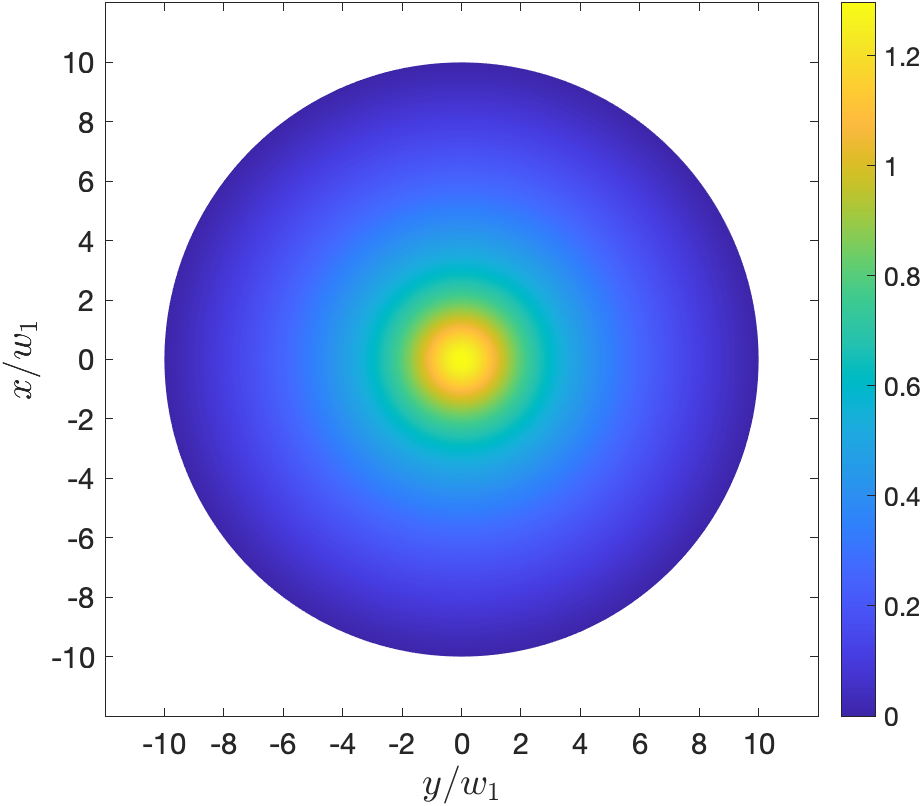}
    \subcaption{$w_x : w_y = 1 : 1$}
    \end{subfigure}
    \hfill
    \begin{subfigure}[b]{.48\textwidth}
    \centering
    \includegraphics[width=\textwidth]{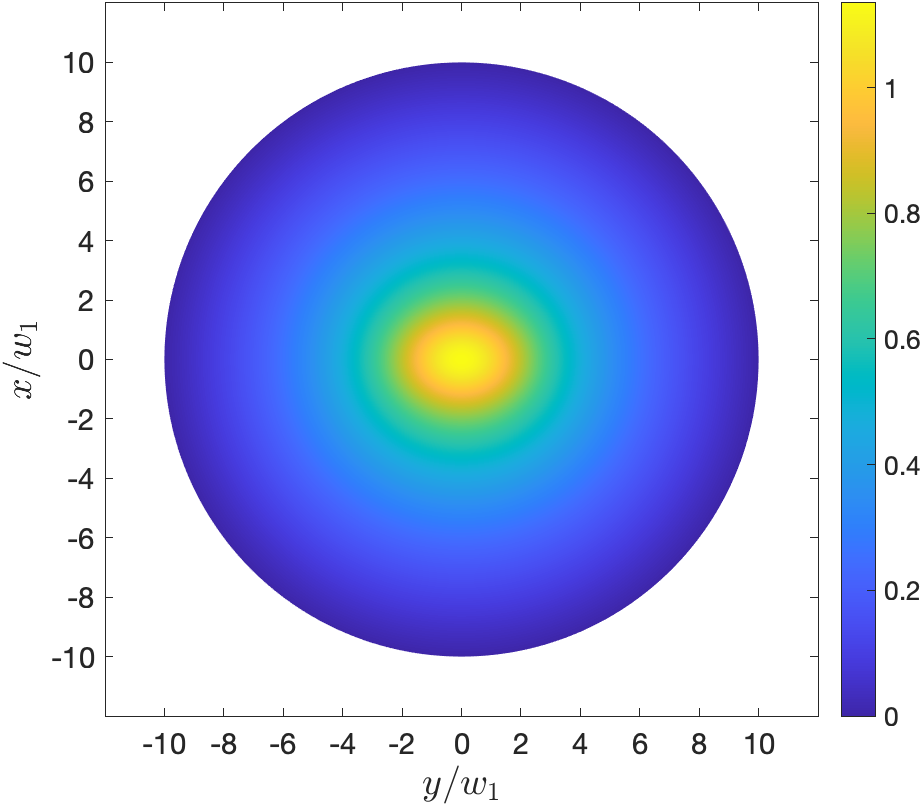}
    \subcaption{$w_x : w_y = 1 : 1.76$}
    \end{subfigure}
    
    \caption{Temperature profiles in the transverse plane of the laser crystal heated by (a) circular Gaussian beam source, (b) elliptical Gaussian beam source, assuming that the total heating power of the two sources are the same.}
    \label{fig:temp_ellip}
\end{figure}

Figure~\ref{fig:temp_ellip} shows the temperature distributions in the transverse plane with circular and elliptical Gaussian beam sources of the same total heating power. In the elliptical case, the ratio of the beam widths is $1 : n$ and we take the refractive index to be $n = 1.76$.

It is found in the numerical simulation that the ratio of 2$^\text{nd}$ order derivatives of temperature at the center of the crystal in the $x-$ and $y-$directions follows
\begin{equation}
    \frac{\del_x^2\, T(0,0)}{\del_y^2\, T(0,0)} = \frac{w_y}{w_x}
\end{equation}
for various ratios of $w_x$ to $w_y$. This rule can be proved for an infinitely large crystal and will be used in the analysis of thermal lens astigmatism in subsection \ref{subsec:thermal_lens_astigmatism}.

\clearpage
% Sec3_thermal_lens.tex
% cannot be compiled alone

\section{Thermal Lens}
\label{sec:thermal_lens}

\subsection{Formula under azimuthal symmetry}
The actual laser beam generated propagates through the crystal near its central axis and the optical path depends on $r$. Under the paraxial approximation, the optical path difference (OPD) is equivalent to that caused by a thin lens of focal length $f_T$ (called “thermal lens” in the text to follow)
\begin{equation}
    \text{OPD}(r) - \text{OPD}(0) = -\frac{r^2}{2f_T}\ .
    \label{eq:lens_opd}
\end{equation}

As the crystal is heated, two effects contribute to the thermal lens: (i) the change in refractive index with temperature and (ii) the change in the length of the crystal with temperature. Assuming that in the range of temperature involved, the changes in refractive index and length are linear to temperature, i.e.\ 
\begin{align}
    \fdd{n}{T} &= \text{const} \\
    \fdd{l}{T} &= 2\alpha_e l_0 = \text{const}
\end{align}
where $\alpha_e$ is the thermal expansion coefficient and $l_0$ is the length at each end section of the crystal over which expansion occurs and is approximately equal to $R$, the radius of the crystal. Therefore, the optical path difference can be written as
\begin{equation}
    \text{OPD}(r) - \text{OPD}(0) = \left[ \fdd{n}{T}\,L + (n-1)\fdd{l}{T} \right](T(r) - T(0))\ .
\end{equation}

To evaluate the temperature difference near the center, we expand $T(r)$ in Taylor series for which its derivatives at $r = 0$ are needed. Eq.~(\ref{eq:temp_r}) gives that
\begin{equation}
    \fdd{T}{r} = - T_p\,\frac{1}{r} \left[ 1- \exp\left( -\frac{r^2}{w_1^2} \right) \right]
    \label{eq:dT_dr}
\end{equation}
where we have used the definition of $T_p$ in Eq.~(\ref{eq:Tp_def}). Differentiating again, 
\begin{equation}
    \fdd{^2 T}{r^2} = - T_p\, \frac{1}{r^2} 
    \left[ -1 + \left( 1 + \frac{2r^2}{w_1^2} \right) 
    \exp\left( -\frac{r^2}{w_1^2} \right) \right]\ .
    \label{eq:dT_dr_2}
\end{equation}
Evaluating Eqs.~(\ref{eq:dT_dr}) and (\ref{eq:dT_dr_2}) at $r = 0$, we have
\begin{align}
    \fdd{T}{r}(r = 0) &= 0\ , 
    \notag \\
    \fdd{^2 T}{r^2}(r = 0) &= - T_p\, \frac{1}{w_1^2} = -\frac{q_0}{2\kappa}\ .
\end{align}

Therefore,
\begin{equation}
    T(r) - T(0) \simeq \frac12\, \del_r^2\, T(0)\, r^2
\end{equation}
and
\begin{equation}
    \text{OPD}(r) - \text{OPD}(0) \simeq \left[ \fdd{n}{T}\,L + 2\alpha_e (n-1)R \right]
    \frac12\, \del_r^2\, T(0)\, r^2 \ .
\end{equation}
Comparing with Eq.~(\ref{eq:lens_opd}), we have that
\begin{equation}
    f_T^{-1} = \left[ \fdd{n}{T}\,L + 2\alpha_e (n-1)R \right](-\del_r^2\, T(0))\ .
    \label{eq:fT_formula_1}
\end{equation}
Thus
\begin{equation}
    f_T = \frac{\pi\kappa w_0^2}{\beta P_0}
    \label{eq:fT_formula_2}
\end{equation}
where
\begin{equation}
    \beta = \fdd{n}{T} + \frac{2\alpha_e (n-1)R}{L}
    \label{eq:beta_def}
\end{equation}
and $P_0 = \pi w_1^2 q_0 L$ is the total heating power dissipated in the laser crystal, with $\kappa$ being the thermal conductivity and $\alpha_e$ the thermal expansion coefficient.

\subsection{Astigmatism}
\label{subsec:thermal_lens_astigmatism}
Though an analytical expression of the temperature distribution can no longer be obtained when the laser crystal is heated by an elliptical Gaussian beam, in order to calculate the thermal lens, we only need the 2$^\text{nd}$ order derivatives of $T$ at the center, as shown by Eq.~(\ref{eq:fT_formula_1}).

It is shown in Appendix~\ref{app:2nd_order_del} that, for $T(x,y)$ following Eqs.~(\ref{eq:temp_xy}) and (\ref{eq:temp_xy_boundary}), when $R \gg w_x, w_y$, we have that
\begin{align}
    \fpp{^2 T}{x^2} (0,0) &= -\frac{q_0}{\kappa}\frac{w_y}{w_x + w_y} \\
    \fpp{^2 T}{y^2} (0,0) &= -\frac{q_0}{\kappa}\frac{w_x}{w_x + w_y} \ .
\end{align}
With the total heating power given by $P_0 = \pi w_x w_y q_0 L$ in the elliptical case, the thermal lenses in the two directions can be written as
\begin{align}
    f_{T,x} &= \frac{\pi\kappa}{\beta P_0}\,w_x(w_x + w_y) 
    = \frac{n+1}{2} \frac{\pi w_0^2\kappa}{\beta P_0}
    \label{eq:fT_formula_x} \\
    f_{T,y} &= \frac{\pi\kappa}{\beta P_0}\,w_y(w_x + w_y) 
    = \frac{n(n+1)}{2} \frac{\pi w_0^2\kappa}{\beta P_0} \ ,
    \label{eq:fT_formula_y}
\end{align}
from which we have that the ratio of thermal lens in the $y$ (horizontal) direction to that in the $x$ (vertical) direction
\begin{equation}
    \frac{f_{T,H}}{f_{T,V}} = \frac{f_{T,y}}{f_{T,x}} = \frac{w_y}{w_x} = n \ .
\end{equation}

\subsection{Enhancement of maximum power through cryogenic cooling}
\label{subsec:enhancement_of_max_power}
The thermal indices of sapphire at room temperature and cryogenic temperatures are listed in Table~\ref{tab:thermal_index}, where the $\dd n/\dd T$ data are from Ref.~\cite{DeFranzo:1993}. The dependence of $\alpha_e$ on temperature below $300\K$ is unclear and we assume that it decreases linearly with temperature as an upper limit estimation.

The thermal conditions influence the optical property of the Ti:sapphire crystal in the form of Eq.~(\ref{eq:fT_formula_2}). It can be seen from the figures in Table~\ref{tab:thermal_index} that if the crystal is cooled from RT to cryogenic temperatures, the thermal conductivity $\kappa$ will increase dramatically and the factor $\dd n/\dd T$ will decrease. Therefore, at the same pump power, a cryogenically cooled crystal will have a much longer thermal lens than a crystal at room temperature. Ideally, if $\kappa/\beta$ increases by some factor, the pump power can be raised by the same factor while keeping the thermal lens at the same value.

\begin{table}[htb]
    \centering
    \large
    \begin{tabular}{|c|c|c|c|}
         \hline
         \gape{\makecell{Temperature \\ $T\ (\text{K})$}} 
         & \makecell{Thermal conductivity \\ $\kappa\ (\text{W}\text{cm}^{-1}\text{K}^{-1})$ }
         & $\D \fdd{n}{T} \ (10^{-6}\K^{-1})$
         & \makecell{Thermal expansion coefficient \\ $\alpha_e\ (10^{-6}\K^{-1})$} \\
         \hline \hline
         \gape{77}     & 10    & 2.8 & 1.4$^{\text{a}}$ \\
         \cline{1-4}
         \gape{100}    & 4.0   & 3.4 & 1.8$^{\text{a}}$ \\
         \cline{1-4}
         \gape{300}    & 0.35  & 13  & 5.5 \\
         \hline
    \end{tabular}
    \caption{Thermal indices of sapphire at room temperature and cryogenic temperatures. $^{\text{a}}$Upper limit estimation assuming that $\alpha_e \propto T$.}
    \label{tab:thermal_index}
\end{table}

By taking typical parameters of the pump laser, we can estimate the relation of thermal lens to the pump power. Firstly, we assume that the pump has a $1/e^2$ radius of $w_0 = 0.5\mm$. Secondly, we assume that all the pump power incident on the laser crystal is absorbed, i.e.
\begin{equation}
    P_{\text{abs}} \approx P_{\text{pump}}\ .
    \label{eq:P_abs=P_pump}
\end{equation}
This is because for a \ch{Ti^3+} doping concentration of $0.1\%$, the small-signal absorption coefficient is $\alpha_0 = 1.5\cm^{-1}$. For a laser crystal with length $L \geq 2\cm$, the transmittance $e^{-\alpha_0 L} \leq 5\%$. Also, the density of \ch{Ti^3+} at $0.1\%$ doping concentration is $n_{\ch{Ti}} = 4.56 \times 10^{19}\cm^{-3}$. From this, we can deduce the absorption cross section
\begin{equation}
    \sigma^{\text{(abs)}} = \frac{\alpha_0}{n_{\ch{Ti}}} = 3.29 \times 10^{-20} \cm^2\ .
\end{equation}
Taking the recovery lifetime for the absorption $\tau_{\text{R}}$ to be of order $10^2\,\text{ps}$, then the saturation intensity of absorption
\begin{equation}
    I_{\text{sat}}^{\text{(abs)}} = \frac{hc/\lambda_{\text{pump}}}{\sigma^{\text{(abs)}} \tau_{\text{R}}} \sim 10^{11}\,\text{W}\,\text{cm}^{-2}
\end{equation}
which is very high ($\lambda_{\text{pump}} = 532\nm$ is the wavelength of the pump). At a pump power of $1000\,\text{W}$, the pump intensity $I_{\text{pump}} \sim 10^5\,\text{W}\,\text{cm}^{-2}$ is still negligible compared to such a saturation intensity. So the absorption coefficient is very close to its small-signal value even at a high pump power. Thirdly, we estimate the fraction of absorbed pump power converted into heat by the wavelengths of the pump and the laser.
\begin{equation}
    \eta_{\text{heat}} = \frac{P_\text{{heat}}}{P_\text{{abs}}} 
    = 1 - \frac{hc/\lambda_{\text{laser}}}{hc/\lambda_{\text{pump}}} \approx 23.8\%
    \label{eq:eta_heat}
\end{equation}
where $\lambda_{\text{laser}} = 698\nm$. Additionally, in the expression of $\beta$ (Eq.~(\ref{eq:beta_def})), we take the refractive index $n = 1.76$ and the ratio $R/L = 0.2$, consistent with a laser crystal with $R = 4\mm$ and $L = 20\mm$. Combining Eqs.~(\ref{eq:P_abs=P_pump}), (\ref{eq:eta_heat}) and (\ref{eq:fT_formula_x}), we have that the thermal lens in the $x-$direction
\begin{equation}
    f_{T,x} \approx \frac{n+1}{2} \frac{\pi w_0^2\kappa}{\beta\, \eta_{\text{heat}}\, P_{\text{pump}}}\ .
\end{equation}

\begin{figure}[htbp]
    \centering
    \includegraphics[width = .7\textwidth]{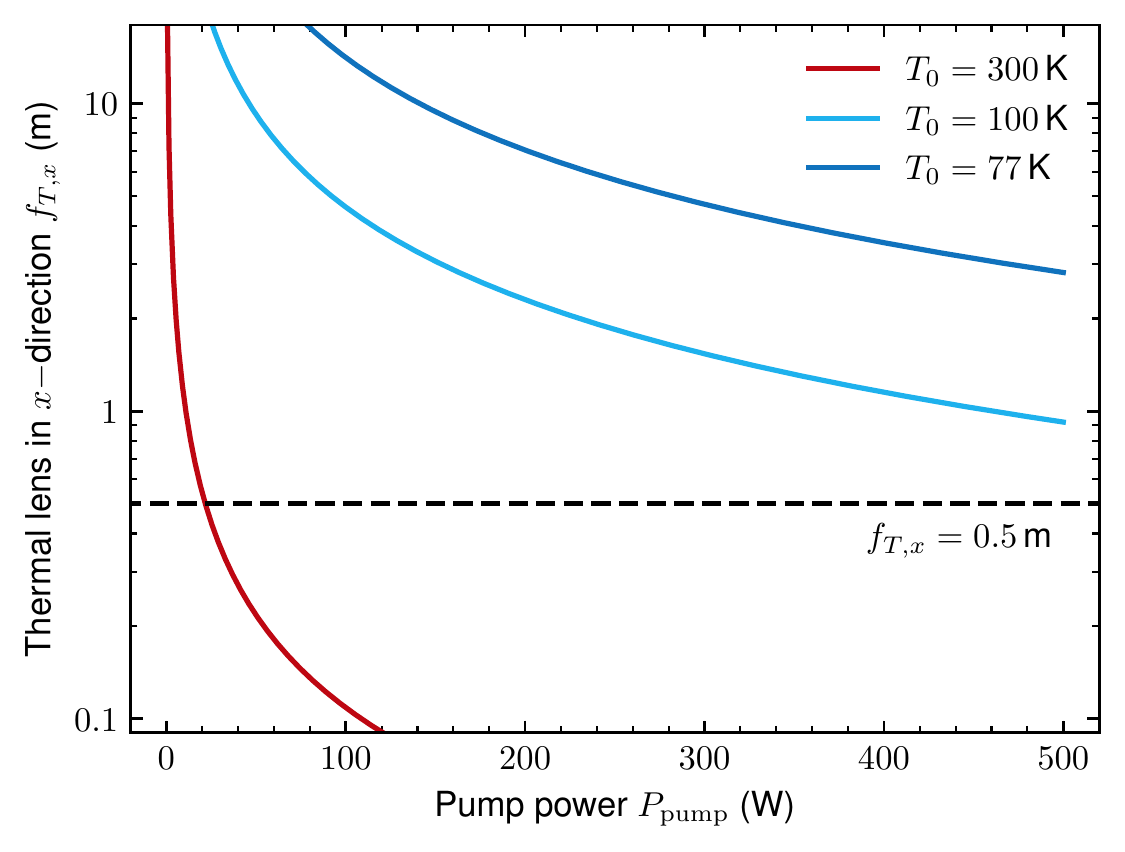}
    \caption{Estimated relation of the focal length of thermal lens in the $x-$direction $f_{T,x}$ to the pump power $P_{\text{pump}}$ when the laser crystal is at $300\K$ (red), $100\K$ (light blue) and $77\K$ (dark blue). The dashed line indicates $f_{T,x} = 0.5\m$ below which the instability of the laser cavity is significant. Note that the vertical axis is logarithmic.}
    \label{fig:fT-P_pump}
\end{figure}

In Figure~\ref{fig:fT-P_pump}, the estimated relation of $f_{T,x}$ to $P_{\text{pump}}$ is evaluated for $T_0 = 300$, $100$ and $77\K$. At the same pump power, the ratio between the focal lengths of thermal lens at the three temperatures is
\begin{equation}
    f_{T,x}(300\K) : f_{T,x}(100\K) : f_{T,x}(77\K) \approx 1 : 40 : 130\ .
\end{equation}
Indicated by the dashed line is $f_{T,x} = 0.5\m$, a benchmark that will be obtained in section~\ref{subsec:stability_against_thermal_lens} below which the instability in the laser cavity will be significant. At $300\K$, $f_{T,x}$ reaches $0.5\m$ at a pump power of about $20\,$W. For an output power of $100\,$W, a pump power of $300\,$W is needed (as given in Appendix~\ref{app:pump_power_for_100W_output}). It is clear that if the laser crystal is cooled to $100\K$ or $77\K$, $f_{T,x}$ is still above $0.5\m$ at such a pump power.

\clearpage
% Sec4_stability_condition.tex
% cannot be compiled alone

\section{Stability Condition of the Laser Cavity}

\subsection{Transfer matrix and stability condition}
\begin{figure}[htbp]
    \centering
    \includegraphics[height=.5\textwidth]{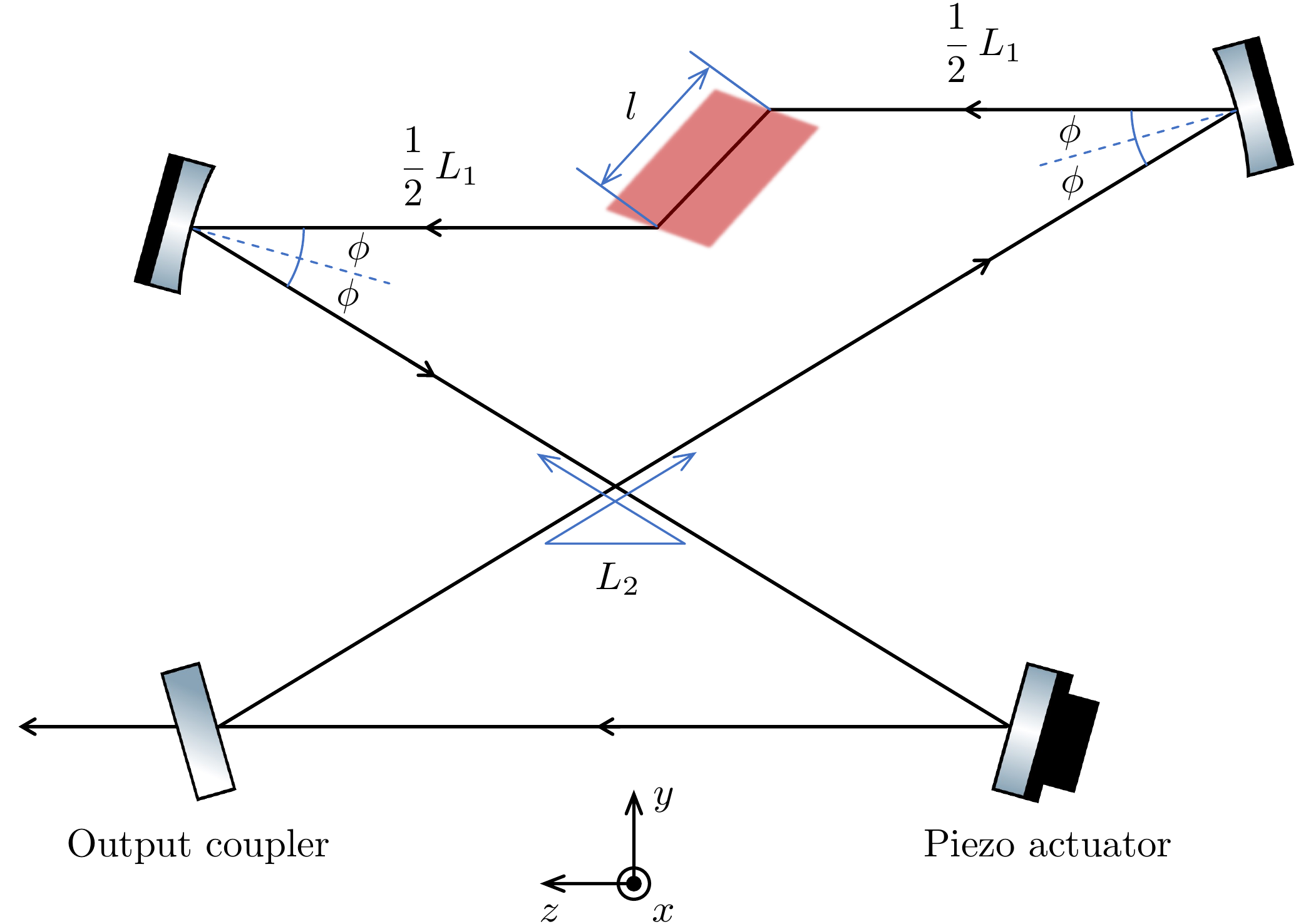}
    \vspace{5mm}
    \caption{Simplified schematic of the Ti:sapphire laser cavity.}
    \label{fig:cavity_full}
\end{figure}

In this section, we calculate the stability condition of the Ti:sapphire laser cavity, expressed in terms of the geometrical parameters. A schematic of the bow-tie cavity is shown in Figure~\ref{fig:cavity_full} where the components that do not affect the stability are omitted. 

As the laser beam propagates in the loop, it undergoes the following processes (starting from the front surface of the crystal):
\begin{enumerate}
    \item Propagating through the crystal (experiencing refraction at two surfaces and thermal lensing within the crystal);
    \item Propagating for a distance of $L_1/2$;
    \item Reflection at mirror of radius of curvature $R$ with incident angle $\phi$;
    \item Propagating for $L_2$;
    \item Reflection at mirror of radius of curvature $R$ with incident angle $\phi$;
    \item Propagating for $L_1/2$.
\end{enumerate}
The overall transfer matrix can be found by multiplying the following transfer matrices for each step.
\begin{enumerate}
    \item Propagation for a distance of $L$
    \begin{equation}
        M_{\text{trans}}(L) = \begin{pmatrix}
        1  &  L \\
        0  &  1
        \end{pmatrix}\ .
    \end{equation}
    \item Reflection from a curved mirror
    \begin{equation}
        M_{\text{mirror}}(R\eff) = \begin{pmatrix}
        1               &  0 \\
        -2/R\eff  &  1
        \end{pmatrix}
    \end{equation}
    where $R\eff$ is the effective radius of curvature.
    \item Propagation in the crystal which has a length $l$, refractive index $n$ and a thermal lens $f_T$ 
    \begin{equation}
        M_{\text{crystal}}(l\eff, f_T) 
        = \begin{pmatrix}
        1               &  l\eff \\
        -1/f_T  &  1
        \end{pmatrix}
    \end{equation}
    where $l\eff$ is the effective length of the crystal.
\end{enumerate}
In these expressions, there are a few parameters that differ for light in the horizontal and  vertical planes which are summarized in Table~\ref{tab:effective_dim}. Because the ratio between the thermal lens in the $y$ and $x$ directions is equal to the refractive index, we refer to $f_{T,x}$ as $f_T$ and $f_{T,y}$ as $n f_T$.

\begin{table}[htbp]
    \centering
    \large
    \begin{tabular}{|c|c|c|c|}
        \hline
         & \gape{\makecell{Effective radius of curvature \\ $R\eff$}}  
         & \makecell{Effective crystal length \\ $l\eff$} 
         & \makecell{Thermal lens \\ $f_T$}   \\
         \hline \hline
    \gape{\makecell{$x-$direction \\ vertical}} & $R/\cos\phi$
    & $l_x = l/n$  &  $f_{T,x} = f_T$    \\
         \hline
    \gape{\makecell{$y-$direction \\ horizontal}} & $R\cos\phi$
    & $l_y = l/n^3$  &  $f_{T,y} = n f_T$   \\
        \hline
    \end{tabular}
    \caption{Parameters that have different values in the vertical ($x$) and horizontal ($y$) directions.}
    \label{tab:effective_dim}
\end{table}

Based on the matrices above, we can define the overall transfer matrix as
\begin{equation}
\begin{split}
    M(L_1,L_2,R\eff,l\eff,f_T) &= M_{\text{trans}}(L_1/2) M_{\text{mirror}}(R\eff) M_{\text{trans}}(L_2)  \\
    &\quad\ M_{\text{mirror}}(R\eff) M_{\text{trans}}(L_1/2) M_{\text{crystal}}(l\eff, f_T) .
\end{split}
\end{equation}
Therefore, the transfer matrices in the $x-$ and $y-$directions can be written as
\begin{align}
    M_x &= M(L_1,\ L_2,\ R/\cos\phi,\ l/n,\ f_T)\ ,
    \label{eq:trans_mat_x}  \\
    M_y &= M(L_1,\ L_2,\ R\cos\phi,\ l/n^3,\ nf_T)\ .
    \label{eq:trans_mat_y}
\end{align}

\subsection{Stability condition}
If we write the overall transfer matrix as
\begin{equation}
    M = \begin{pmatrix}
    A & B \\
    C & D
    \end{pmatrix},
\end{equation}
then the complex radius of the Gaussian beam mode is given by
\begin{equation}
    q = \frac{1}{2C}\left[ (A-D) \pm \sqrt{(A-D)^2 + 4BC} \right]
\end{equation}
and the condition for the laser cavity to be stable is
\begin{equation}
    \Delta = (A-D)^2 + 4BC < 0
\end{equation}
(specific derivation is given in Appendix~\ref{app:stab_cond}).

Since
\begin{equation}
    q = z - z_0 + \ii \zR\ ,
\end{equation}
we can define the squared Rayleigh range as
\begin{equation}
    \zR^2 = -\frac{1}{4C^2}\left[ (A-D)^2 + 4BC \right]\ .
\end{equation}
In this way, the condition for the laser cavity being stable is equivalent to
\begin{equation}
    \zR^2 > 0\ .
\end{equation}

\subsection{Astigmatism compensation}
With the transfer matrices in the $x-$ and $y-$directions given by Eqs.~(\ref{eq:trans_mat_x}) and (\ref{eq:trans_mat_y}), the squared Rayleigh ranges in the two directions can be written as functions of 6 parameters
\begin{align}
    z^2_{\text{R},x} &= \zeta_x(L_1, L_2, R, \phi, l, f_T) \\
    z^2_{\text{R},y} &= \zeta_y(L_1, L_2, R, \phi, l, f_T)\ .
\end{align}
In general, $z^2_{\text{R},x}$ and $z^2_{\text{R},y}$ are not the same and the range of parameters for $\zR^2 > 0$ are different.

\begin{figure}[tp]
    \begin{subfigure}[b]{.48\textwidth}
    \centering
    \includegraphics[width=\textwidth]{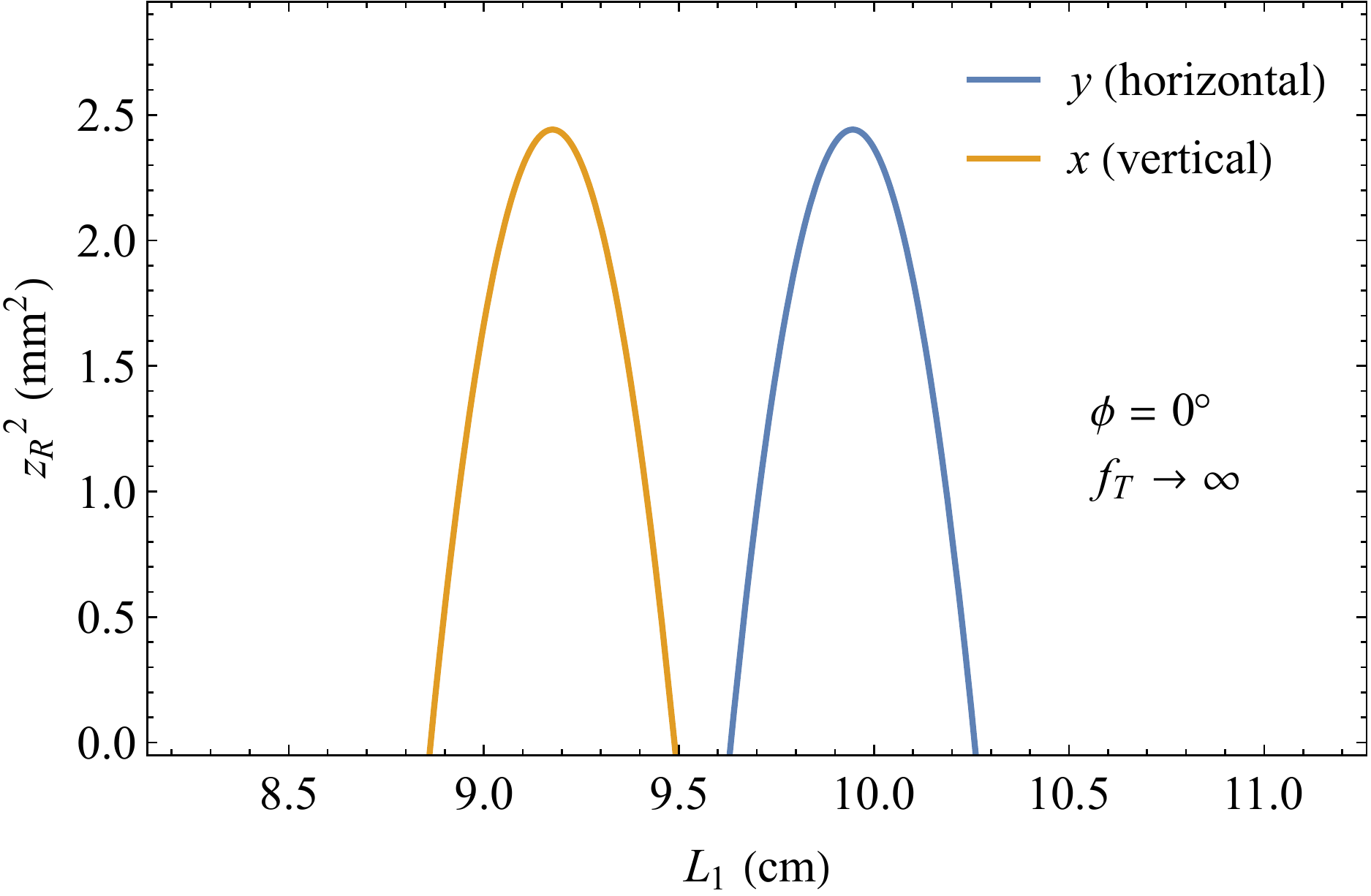}
    \subcaption{}
    \label{subfig:R=10_zR-L1_a}
    \end{subfigure}
    \hfill
    \begin{subfigure}[b]{.48\textwidth}
    \centering
    \includegraphics[width=\textwidth]{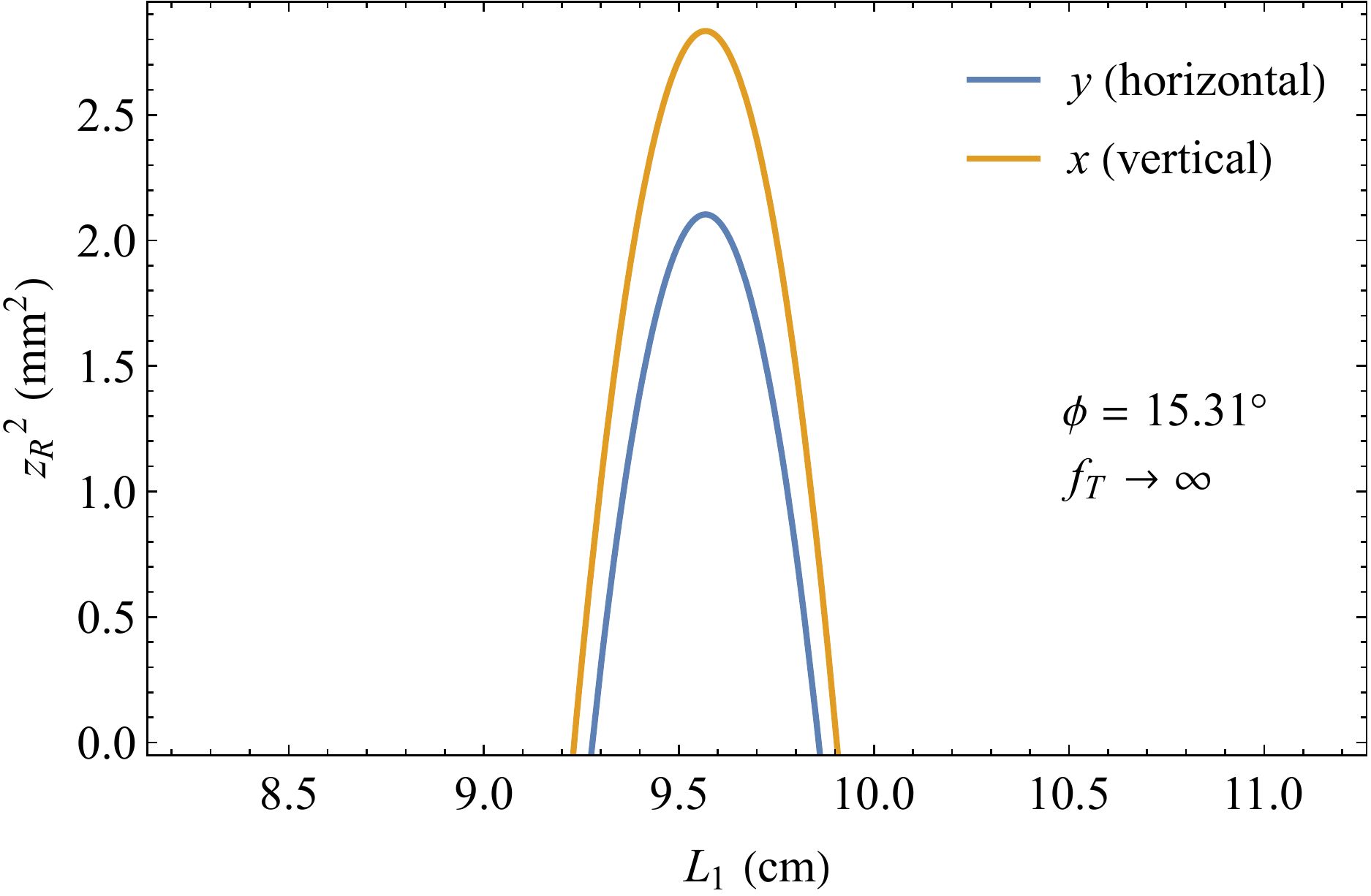}
    \subcaption{}
    \label{subfig:R=10_zR-L1_b}
    \end{subfigure}
    
    \caption{The relation of $\zR^2$ to $L_1$ in the $x$ (vertical) and $y$ (horizontal) directions, for the incident angle on the spherical mirror $\phi$ being (a) $0^{\circ}$ where the working regions do not overlap and (b) $15.31^{\circ}$ which gives astigmatism compensation.}
    \label{fig:R=10_zR-L1}
\end{figure}

For example, for a particular bow-tie cavity with parameters given in Ref.~\cite{Lin:1995},
\begin{align*}
    L_2 &= 170 \cm \\
    R &= 10 \cm \\
    l &= 2 \cm\ ,
\end{align*}
if we do not consider the thermal lens (i.e.\ taking $f_T \rightarrow \infty$), then at $\phi = 0^{\circ}$, the ranges of $L_1$ for the cavity to be stable in the $x-$ and $y-$directions do not overlap and the cavity can never function properly, as shown in Figure~\ref{subfig:R=10_zR-L1_a}. Instead, $\phi$ needs to be chosen carefully so that the centers of the two ranges coincide and this is found at $\phi = 15.31^{\circ}$, as shown in Figure~\ref{subfig:R=10_zR-L1_b}.

We refer to the range of $L_1$ where $\zR^2 > 0$ as the working region. In general, the $\zR^2$ vs.\ $L_1$ curve within the working region is approximately parabolic with a single maximum. We refer to the value of $L_1$ where $\zR^2$ takes this maximum as the optimal working region. This is mathematically defined by
\begin{align}
    L_{1,x} \quad \text{is such that} \quad
    \fpp{\zeta_x}{L_1}(L_1 = L_{1,x}, L_2, R, \phi, l, f_T) = 0
    \label{eq:L_1x_def} \\
    L_{1,y} \quad \text{is such that} \quad
    \fpp{\zeta_y}{L_1}(L_1 = L_{1,y}, L_2, R, \phi, l, f_T) = 0
    \label{eq:L_1y_def}
\end{align}
for the $x-$ and $y-$directions. We want the laser cavity to operate near the optimal working region so that the influence of tiny changes in $L_1$ upon the stability of the laser is minimized. 

The optimal working regions of the $x-$ and $y-$directions are also different in general. To compensate the astigmatism is to choose a proper set of parameters such that $L_{1,x}$ and $L_{1,y}$ coincide, and we choose to do so at zero power (i.e. when $f_T \rightarrow \infty$). This is given by
\begin{gather}
    L_{1,x} = L_{1,y} = L_1 \notag \\
    \Leftrightarrow \quad
    \fpp{\zeta_x}{L_1}(L_1, L_2, R, \phi, l, \infty) =
    \fpp{\zeta_y}{L_1}(L_1, L_2, R, \phi, l, \infty) = 0\ .
    \label{eq:astigma_compen}
\end{gather}
Eq.~(\ref{eq:astigma_compen}) imposes two restrictions. If we take the crystal length $l$ and the second path length $L_2$ as given parameters, then we can determine the three remaining parameters $L_1$, $R$ and $\phi$ based on one of them. In practice, $L_1$ and $\phi$ can be adjusted precisely but only mirrors with particular $R$ are available. Therefore, we calculate $L_1$ and $\phi$ from $R$. Though the expression of $\zR$ is complicated, Eq.~(\ref{eq:astigma_compen}) can be simplified into 2 polynomials in terms of $L_1$ and $\cos\phi$ which can be solved numerically.

To sum up, at no thermal lens, $L_1$ and $\phi$ can be determined from given $L_2$, $R$ and $l$ according to Eq.~(\ref{eq:astigma_compen}) such that the optimal working regions of the $x-$ and $y-$directions are coincident.

\clearpage

\begin{figure}[bp]
    \centering
    \includegraphics[width = .5\textwidth]{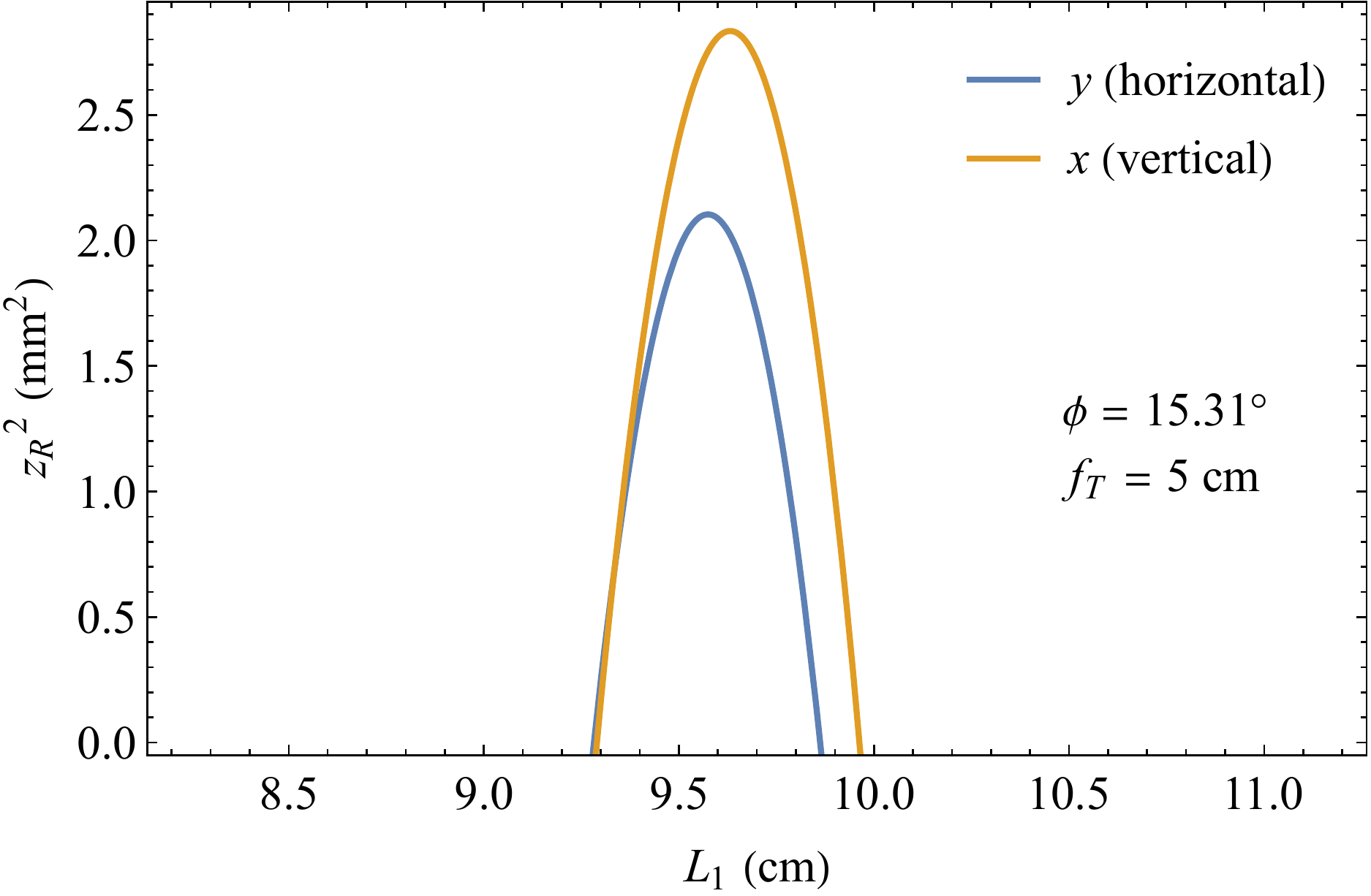}
    \caption{The relations of $\zR^2$ to $L_1$ under the same conditions as in Figure~\ref{subfig:R=10_zR-L1_b} except that $f_T$ is now $5\cm$. Astigmatism compensation is broken by the thermal lens.}
    \label{fig:R=10_fT_distort}
\end{figure}

\clearpage
\subsection{Optimal parameters}
Using the protocol introduced in the previous subsection, we can obtain several sets of cavity parameters which give astigmatism compensation. These are chosen for $R = 10$, $20$ and $30\cm$ respectively, listed in Table~\ref{tab:optimal_parameter}.

\begin{table}[h!]
    \centering
    \large
    \begin{tabular}{|c|c|c|c|c|}
        \hline
        \gape{$L_1$ (cm)} & $L_2$ (cm) & $R$ (cm) & $\phi$ ($^{\circ}$) & $l$ (cm) \\
        \hline \hline
        \gape{9.568} & 170.0 & 10.00 & 15.31 & 2.00  \\
        \hline
        \gape{19.975} & 240.0 & 20.00 & 11.96 & 2.50  \\
        \hline
        \gape{31.207} & 240.0 & 30.00 & 9.53  & 2.50  \\
        \hline
    \end{tabular}
    \caption{Three sets of cavity parameters with astigmatism compensation where the radii of curvature of the spherical mirror $R$ are $10$, $20$ and $30\cm$ respectively.}
    \label{tab:optimal_parameter}
\end{table}

\subsection{Stability against thermal lens}
\label{subsec:stability_against_thermal_lens}
With increasing effect of thermal lens ($f_T$ decreasing from $\infty$), the optimal working regions are shifted and the $\zR^2$ vs.\ $L_1$ curves are distorted. Figure~\ref{fig:R=10_fT_distort} shows the $\zR^2$ vs.\ $L_1$ curves for the astigmatism compensated design at a strong thermal lens of $f_T = 5\cm$.

\begin{figure}[htbp]
    \begin{subfigure}[b]{.49\textwidth}
    \centering
    \includegraphics[width=\textwidth]{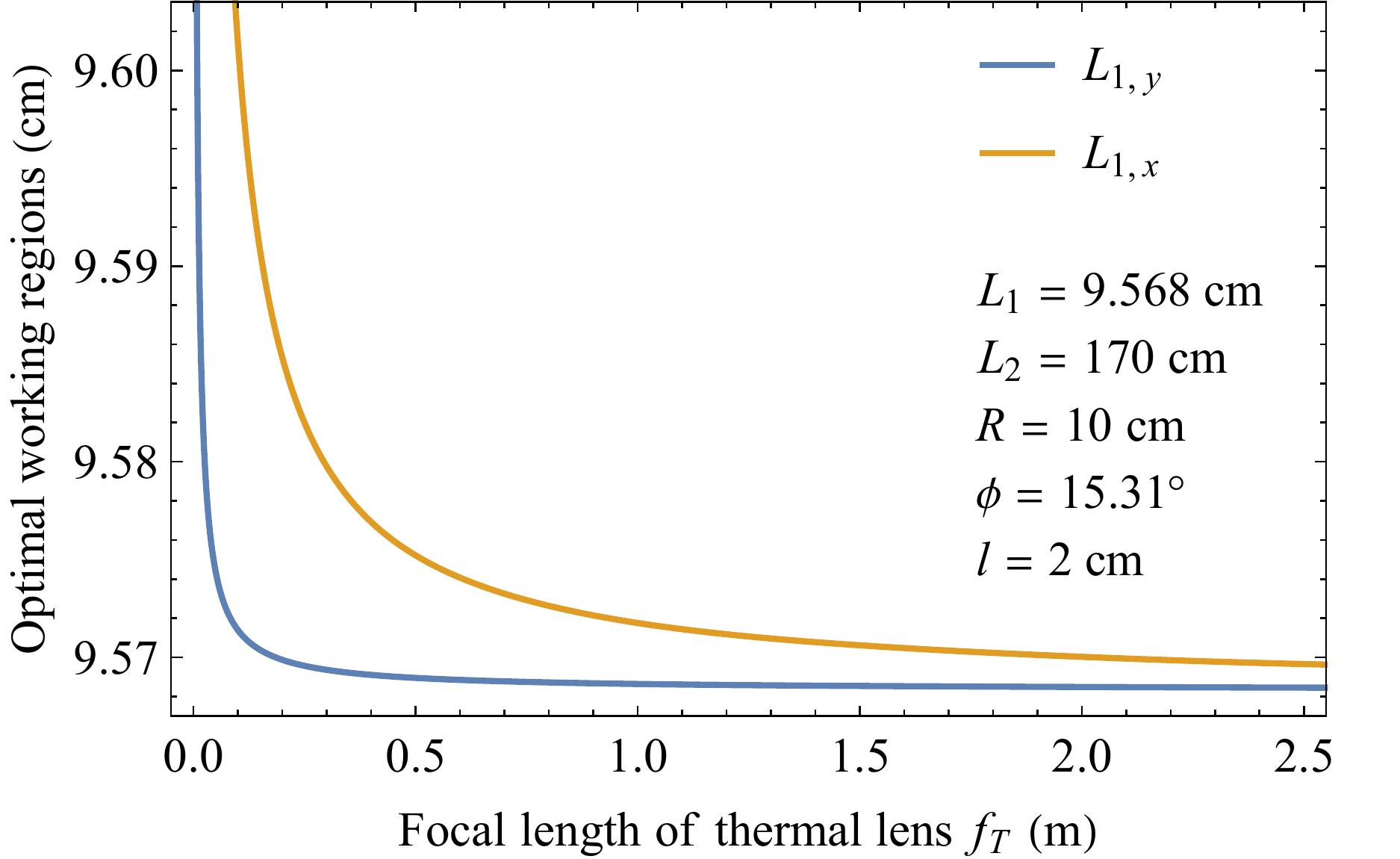}
    \subcaption{$R = 10\cm$, $l = 2.0\cm$}
    \label{subfig:L1-fT_a}
    \end{subfigure}
    \hfill
    \begin{subfigure}[b]{.49\textwidth}
    \centering
    \includegraphics[width=\textwidth]{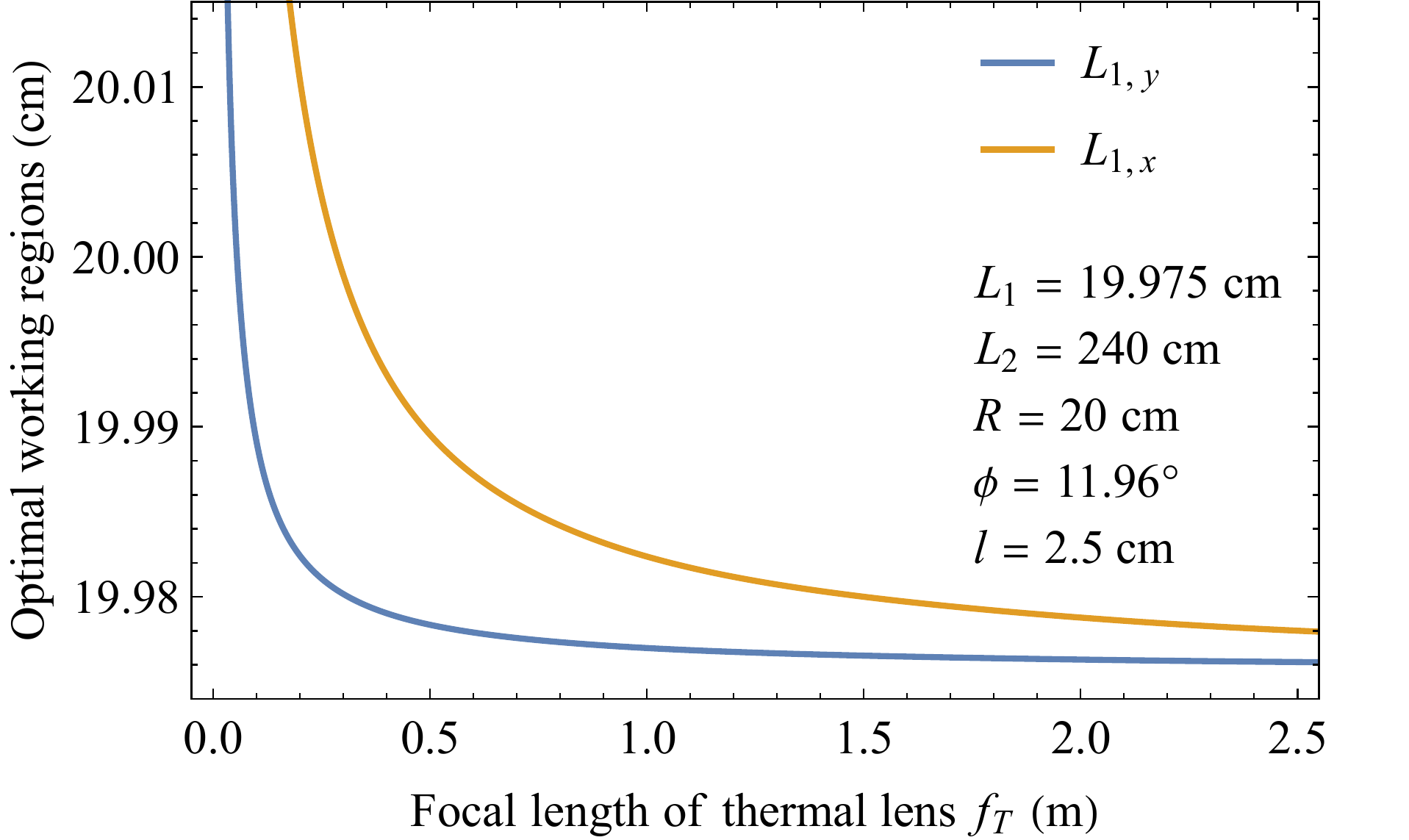}
    \subcaption{$R = 20\cm$, $l = 2.5\cm$}
    \label{subfig:L1-fT_b}
    \end{subfigure}
    \vspace{5mm}

    %\hspace{0.5mm}
    \begin{subfigure}[b]{.49\textwidth}
    \centering
    \includegraphics[width=\textwidth]{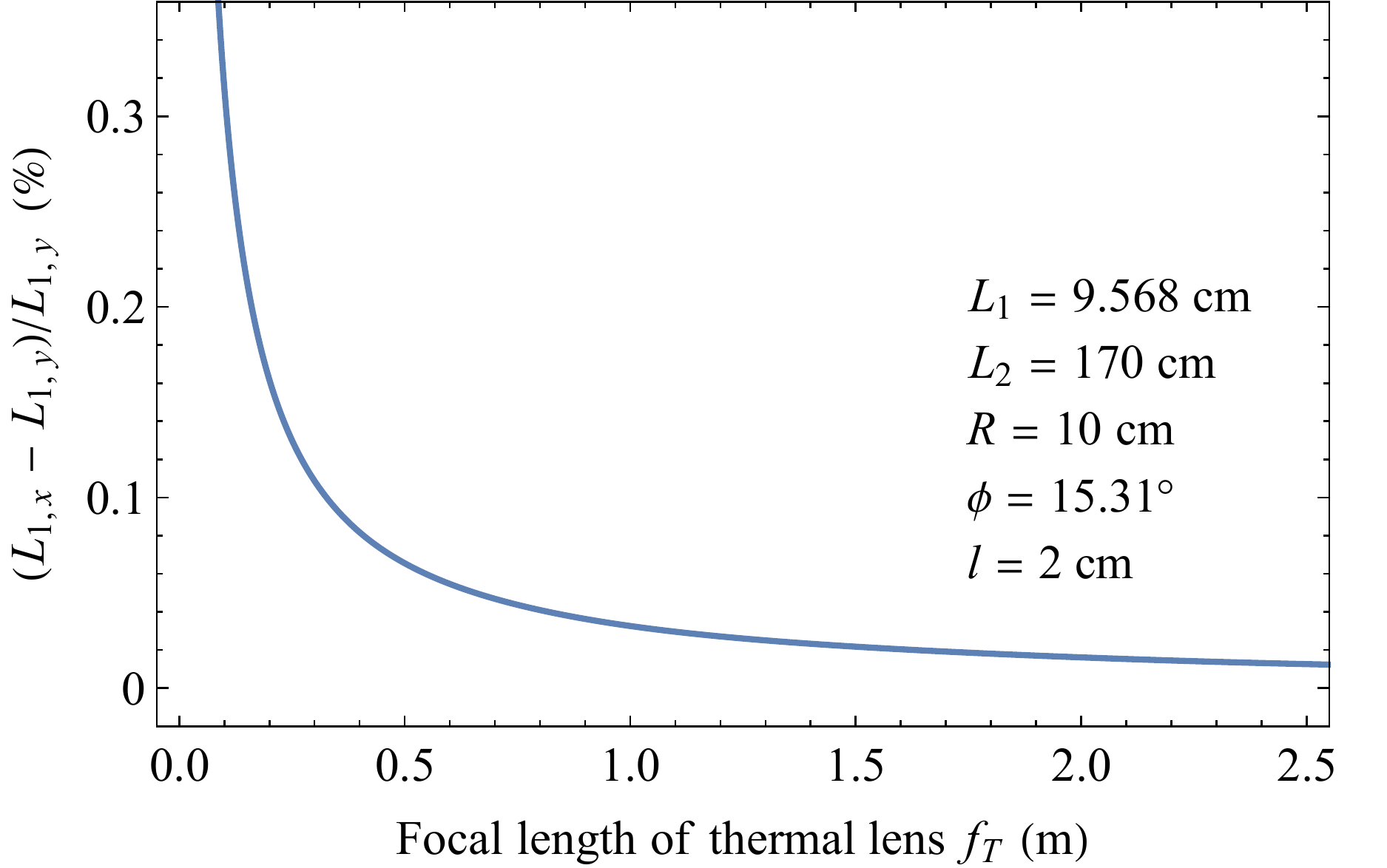}
    \subcaption{$R = 10\cm$, $l = 2.0\cm$}
    \label{subfig:L1-fT_c}
    \end{subfigure}
    \hfill
    \begin{subfigure}[b]{.49\textwidth}
    \centering
    \includegraphics[width=\textwidth]{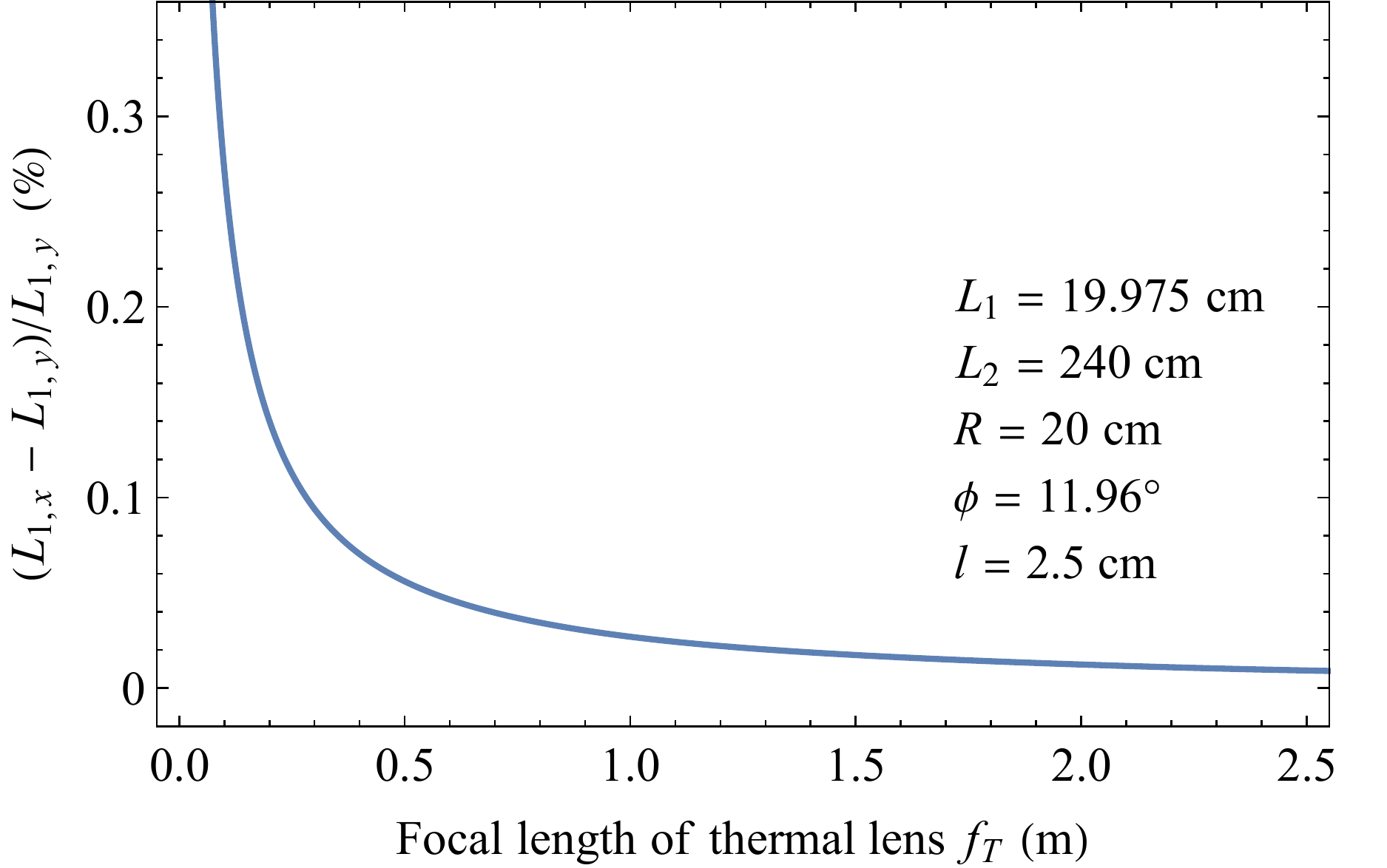}
    \subcaption{$R = 20\cm$, $l = 2.5\cm$}
    \label{subfig:L1-fT_d}
    \end{subfigure}
    
    \caption{(a), (b) Relation of optimal working regions in the $x$ (vertical) direction $L_{1,x}$ and in the $y$ (horizontal) direction $L_{1,y}$, as defined in Eqs.~(\ref{eq:L_1x_def}) and (\ref{eq:L_1y_def}), to the focal length of thermal lens $f_T$ for two astigmatism compensated designs with $R = 10$ and $20\cm$. (c), (d) Relative difference between $L_{1,x}$ and $L_{1,y}$ in (a), (b) as functions of $f_T$. The laser cavity parameters are listed at the right of each figure where $L_1\,:$ first path length, $L_2\,:$ second path length, $R\,:$ spherical mirror radius of curvature, $\phi\,:$ incident angle on the spherical mirror and $l\,:$ laser crystal length. $L_1$ and $\phi$ are chosen such that $L_{1,x} = L_{1,y} = L_1$ when $f_T \rightarrow \infty$. Note that (a), (c) characterize the same design and (b), (d) characterize the same design. For both designs, as $f_T$ decreases, $L_{1,x}$ and $L_{1,y}$ increase and their relative difference also increases.}
    \label{fig:L1-fT}
\end{figure}

The response to $f_T$ for optimized designs with different $R$ are similar but differ in magnitude. To characterize the influence of thermal lens accurately, we plot the optimal working regions of the $x-$ and $y-$directions, i.e.\ $L_{1,x}$ and $L_{1,y}$ as defined in Eqs.~(\ref{eq:L_1x_def}) and (\ref{eq:L_1y_def}), and their relative difference $(L_{1,x} - L_{1,y})/L_{1,y}$ against $f_T$ for the first two sets of parameters in Table~\ref{tab:optimal_parameter}, shown in Figure~\ref{fig:L1-fT}. In these graphs, optimal working regions are evaluated for $f_T$ below $2.5\,\text{m}$. Comparing the behavior of the two designs, we can find that the one with larger $R$ (as shown in Figs.~\ref{subfig:L1-fT_b} and \ref{subfig:L1-fT_d}) has both a smaller percentage increase in optimal working regions and a smaller relative difference between optimal working regions in the two directions. 

Besides, since $f_T$ enters the expressions of transfer matrix and Rayleigh range in the form of $1/f_T$, all the curves in Figure~\ref{fig:L1-fT} have the shape of the inverse scale function. It can be estimated that in order for the relative difference between the optical working regions in the $x-$ and $y-$directions to be below $0.1\%$, that is $(L_{1,x} - L_{1,y})/L_{1,y} \leq 0.1\%$ in Figs.~\ref{subfig:L1-fT_c} and \ref{subfig:L1-fT_d}, the focal length of thermal lens needs to be
\begin{equation}
    f_T \gtrsim 0.5\,\text{m}\ .
    \label{eq:fT>0.5m}
\end{equation}
This is how the benchmark we used in section~\ref{subsec:enhancement_of_max_power} is derived. As shown in Figure~\ref{fig:fT-P_pump}, Eq.~(\ref{eq:fT>0.5m}) is satisfied for pump power up to $500\,$W, given that the laser crystal is cooled to temperatures of $100\K$ or lower. Therefore, we can expect the instability of the laser cavity to be insignificant if the laser crystal is cryogenically cooled.

\clearpage
% Sec5_conclusion.tex
% cannot be compiled alone

\section{Conclusion}
In this study, we analyzed the practicality of enhancing the power of Ti:sapphire laser by cryogenically cooling the laser crystal. We obtained analytical expressions of the focal length of thermal lens for the pump inside the laser crystal being a circular Gaussian beam or an elliptical Gaussian beam. In order for the relative difference between the optimal laser cavity path lengths in the horizontal and vertical directions to be less than $0.1\%$, the focal length of thermal lens should be longer than $0.5\m$. Such a requirement can be met if the laser crystal is cooled to temperatures below $100\K$ while being pumped at a power that is enough to produce $100\,$W output. In conclusion, it is very likely that an output power of $100\,$W can be achieved on a single Ti:sapphire laser by cryogenic cooling of the laser crystal.

\bibliographystyle{unsrt}
\bibliography{Reference}

\clearpage
\appendix
% Sec6_appendix.tex
% cannot be compiled alone

%\appendix

\begin{appendices}

%\section{Appendix}
%\renewcommand{\thesubsection}{\Alph{subsection}}
\renewcommand{\thefigure}{\thesection\arabic{figure}}
\renewcommand{\theequation}{\thesection\arabic{equation}}
\numberwithin{figure}{section}
\numberwithin{equation}{section}

\section{Relationship of thermal conductivity of sapphire to temperature}
\label{app:thermal_index}
\begin{figure}[h!]
    \centering
    \includegraphics[width=.96\textwidth]{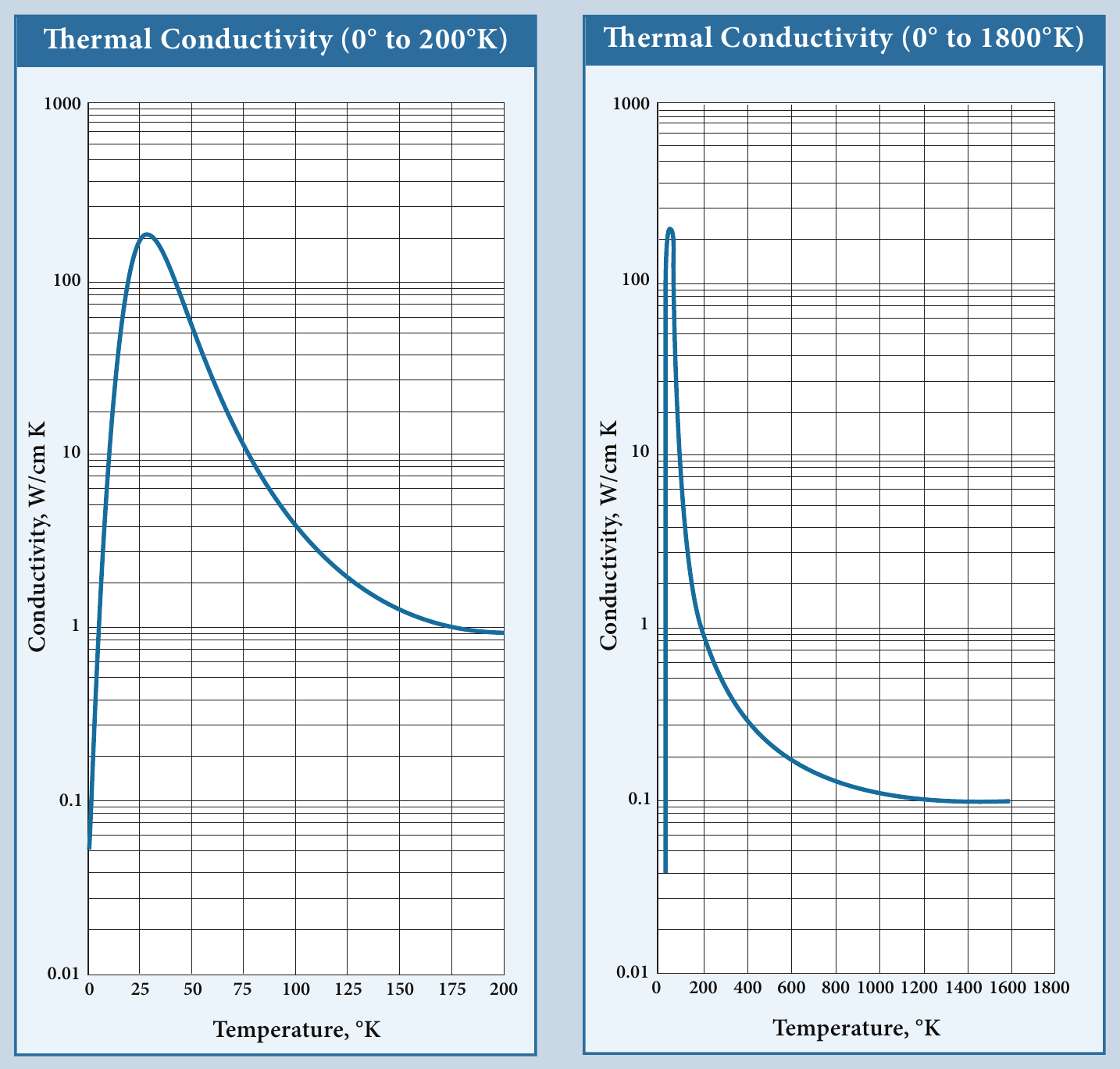}
    \caption{Thermal conductivity of sapphire (\ch{Al2O3}) vs.\ temperature. Note that the vertical axis is logarithmic. After~\cite{rayotek}.}
    \label{fig:sapphire_thermal_index}
\end{figure}

\clearpage
\section{Detailed Derivations}

\subsection{Second order derivatives of temperature at the origin}
\label{app:2nd_order_del}
In Eqs.~(\ref{eq:u_xy}) and (\ref{eq:u_xy_boundary}), when the radius of the laser crystal is much larger than the widths of the beam, i.e.\ $R \gg w_x,w_y$, the boundary condition can be approximated by
\begin{equation*}
    u \rightarrow 0 \qquad \text{at } x^2 + y^2 \rightarrow \infty\ .
\end{equation*}

Under such circumstance, $u(x,y)$ can in principle be calculated through Fourier transform. While the analytical expression of $u(x,y)$ is hard to get, its second order derivatives at the origin can be obtained. The field $u(x,y)$ and its Fourier transform $\hat{u}(k_x, k_y)$ are linked by
\begin{equation*}
    u(x,y) = \frac{1}{2\pi} \int_{-\infty}^{\infty} \int_{-\infty}^{\infty} 
    \hat{u}(k_x, k_y) \exp[\ii(k_x x + k_y y)] \,\dd k_x\,\dd k_y\ .
\end{equation*}
Therefore,
\begin{align}
    \del_x^2\, u &= \frac{1}{2\pi} \iint (-k_x^2)\,
    \hat{u} \exp[\ii(k_x x + k_y y)] \,\dd k_x\,\dd k_y
    \label{app_eq:u_xx} \\
    \del_y^2\, u &= \frac{1}{2\pi} \iint (-k_y^2)\,
    \hat{u} \exp[\ii(k_x x + k_y y)] \,\dd k_x\,\dd k_y
    \label{app_eq:u_yy}
\end{align}
where the integration limits on $(-\infty,\infty)$ are dropped. We also have that the Fourier transform of the source function
\begin{equation*}
    \hat{q}(k_x, k_y) = 
    \frac12\, w_x w_y \exp\left[ -\frac14 \left( w_x^2 k_x^2 + w_y^2 k_y^2 \right) \right]\ .
\end{equation*}
So $\hat{u}(k_x, k_y)$ follows the equation
\begin{equation*}
    (k_x^2 + k_y^2)\,\hat{u} = \frac12\, w_x w_y \exp\left[ -\frac14 \left( w_x^2 k_x^2 + w_y^2 k_y^2 \right) \right]
\end{equation*}
of which the solution is
\begin{equation*}
    \hat{u} = \frac12\, w_x w_y\, \frac{1}{k_x^2 + k_y^2} \exp\left[ -\frac14 \left( w_x^2 k_x^2 + w_y^2 k_y^2 \right) \right]\ .
    \label{eq:}
\end{equation*}
Therefore, $u(x,y)$ can formally be written as
\begin{equation*}
    u(x,y) = \frac{1}{4\pi}\, w_x w_y \iint 
    \frac{1}{k_x^2 + k_y^2} \exp\left[ -\frac14 \left( w_x^2 k_x^2 + w_y^2 k_y^2 \right) \right] 
    \exp[\ii(k_x x + k_y y)] \,\dd k_x\,\dd k_y
\end{equation*}
but the integral can hardly be evaluated.

\begin{figure}[ht]
    \centering
    \includegraphics[width = .99\textwidth]{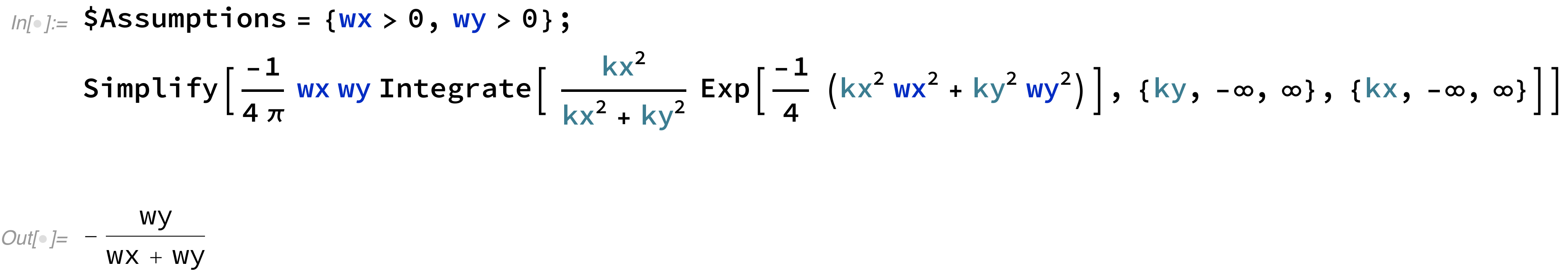}
    \caption{{\small MATHEMATICA}'s evaluation of Eq.~(\ref{app_eq:u_xx_evaluate}).}
    \label{fig:2nd_order_del}
\end{figure}

Substituting into Eq.~(\ref{app_eq:u_xx}) and evaluating at $(x,y) = (0,0)$,
\begin{equation}
\begin{split}
    \fpp{^2 u}{x^2}(0,0) &= -\frac{1}{4\pi}\,w_x w_y
    \iint \frac{k_x^2}{k_x^2 + k_y^2} 
    \exp\left[ -\frac14 \left( w_x^2 k_x^2 + w_y^2 k_y^2 \right) \right]
    \,\dd k_x \,\dd k_y  \\
    &= -\frac{w_y}{w_x + w_y}
    \label{app_eq:u_xx_evaluate}
\end{split}
\end{equation}
where the second equality is given by {\small MATHEMATICA} as shown in Figure~\ref{fig:2nd_order_del}. Swapping $w_x \leftrightarrow w_y$, we can get
\begin{equation*}
    \fpp{^2 u}{y^2}(0,0) = -\frac{w_x}{w_x + w_y}
\end{equation*}
which can also be obtained through the relation that
\begin{equation*}
    \fpp{^2 u}{x^2}(0,0) + \fpp{^2 u}{y^2}(0,0) = -1
\end{equation*}
which is Eq.~(\ref{eq:u_xy}) evaluated at $(x,y) = (0,0)$.

According to the manner of variable changing in Eq.~(\ref{eq:u_to_T}), we thus have
\begin{align}
    \fpp{^2 T}{x^2} (0,0) &= -\frac{q_0}{\kappa}\frac{w_y}{w_x + w_y} \\
    \fpp{^2 T}{y^2} (0,0) &= -\frac{q_0}{\kappa}\frac{w_x}{w_x + w_y} \ .
\end{align}

\clearpage
\subsection{Stability condition of the laser cavity}
\label{app:stab_cond}
As for a Gaussian beam, the transfer matrix acts on the vector
\begin{equation*}
    \begin{pmatrix}
    q \\
    1
    \end{pmatrix}
\end{equation*}
where $q$ is the complex radius. After one round-trip, the complex radius should be unchanged. Therefore
\begin{align}
    &\begin{pmatrix}  q \\ 1  \end{pmatrix}
    = \mathcal{UF}
    \begin{pmatrix}  A & B  \\  C & D  \end{pmatrix}
    \begin{pmatrix}  q \\ 1  \end{pmatrix}\ . \\
\intertext{where $\mathcal{UF}$ is an uninteresting factor,}
    \Rightarrow \quad  &q = \frac{Aq + B}{Cq + D} 
    \notag \\
    \Rightarrow \quad  &C q^2 + (D-A) q - B = 0 
    \notag \\
    \Rightarrow \quad  &q = \frac{1}{2C}\left[ (A-D) \pm \sqrt{(A-D)^2 + 4BC} \right]\ .
\end{align}
In order for the mode to be a Gaussian beam, $q$ must be complex, therefore the discriminant must be negative, i.e.\ 
\begin{equation}
    \Delta = (A-D)^2 + 4BC < 0\ .
\end{equation}

\clearpage
\subsection[Pump power needed for 100 W output]{Pump power needed for 100\,W output}
\label{app:pump_power_for_100W_output}
\begin{figure}[hbp]
    \centering
    \includegraphics[width = .8\textwidth]{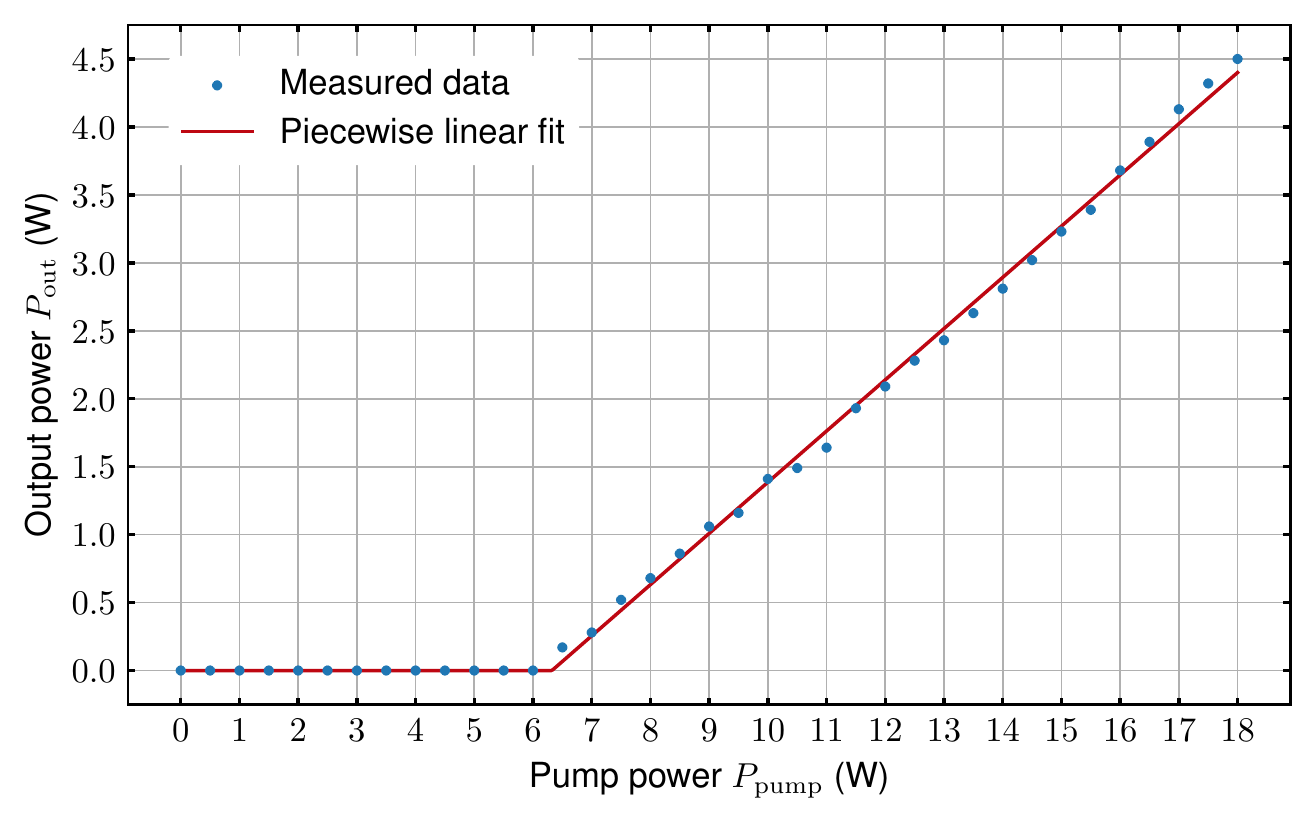}
    \caption{Measured relation of output power to pump power for a M Squared SolsTiS laser operating at $698\nm$ and room temperature. Red line shows piecewise linear fit as defined by 
    Eqs.~(\ref{app_eq:P_out_above_thresh}) and (\ref{app_eq:P_out_below_thresh}).}
    \label{fig:P_out-P_pump_measured}
\end{figure}

Figure~\ref{fig:P_out-P_pump_measured} shows the measured relation of output power to pump power for a M Squared SolsTiS laser operating at $698\nm$ and room temperature. This relation can be fitted with a piecewise linear function
\begin{numcases}{P_{\text{out}} = }
        k(P_{\text{pump}} - P_{\text{thresh}}) \quad & $P_{\text{pump}} \geq P_{\text{thresh}}$
        \label{app_eq:P_out_above_thresh} \\
        \notag \\
        0 \quad & $P_{\text{pump}} < P_{\text{thresh}}$
        \label{app_eq:P_out_below_thresh}
\end{numcases}
where $P_{\text{thresh}}$ is the threshold pump power. The fitting parameters are $k = 0.377$ and $P_{\text{thresh}} = 6.32\,$W. The good linear relation between output power and pump power above lasing threshold means that the laser cavity is stable and that the effect of thermal lens is insignificant.

As we showed in Figure~\ref{fig:fT-P_pump}, if the laser crystal is cooled to cryogenic temperatures, the thermal lens will not break the stability of the laser cavity up to $500\,$W pump power. Under such circumstance, we can expect the output power to continue growing with the pump power linearly according to Eq.~(\ref{app_eq:P_out_above_thresh}) and $P_{\text{out}}$ will reach $100\,$W when $P_{\text{pump}} = 249\,$W. So, in round numbers, $300\,$W pump power is enough to produce $100\,$W output.

\clearpage
\section{Early work on cooled Ti:sapphire lasers in the 1990s}
The advantages of cryocooling have been recognized since the early days of Ti:Sa lasers. A paper from 1991~\cite{Schulz:1991} states that “At 93K, sapphire (\ch{Al2O3}) has much better thermal properties than at room temperature. By taking advantage of the improved thermal properties, a Ti:\ch{Al2O3} laser is operated with 350\,W of output power under thermally steady-state conditions. Model calculations of thermal lensing and higher order thermo-optic aberrations indicate that the output power capability is 200 times larger at 77\,K than at 300\,K”, where ‘thermally steady-state conditions’ means experimental measurements using so-called macro pulses of 170 microseconds (longer than the thermal response time). This quasi-CW operation demonstrated that high-power operation would be sustainable continuously (actual ‘steady-state’) if a high-power pump laser was available. Good thermal contact between the sapphire and the surrounding copper block, attached to a dewar, was achieved by a thin sheet of indium squeezed between them (a standard cryogenic technique, and other experimental aspects are also still relevant for the more powerful pump laser systems now available). A $43\,\mathrm{W}$, CW Ti:sapphire laser was reported at the CLEO conference in 1991~\cite{Erbert:1991}.

\clearpage
\section{Related industrial companies and project}
\label{app:companies_and_project}

\subsection{Kapteyn-Murane Labs (KMLabs)}
\label{app:kmlabs}
This spin-out from University of Colorado, Boulder\footnote{\url{www.kmlabs.com}} manufactures short pulse systems with cryocooled Ti:sapphire laser oscillators (femtosecond pulses and products to generate vacuum ultraviolet (VUV) and extreme ultraviolet (EUV or XUV) radiation from them). This company produces cryocooled modules but the Ti:Sa crystals are generally shorter (to reduce dispersion) than optimal for CW operation, however they “see no fundamental issue in making a $> 30$\,W at 698\,nm Ti:Sa CW laser” but it would require R\&D to implement this. KMLabs has delivered custom CW systems but the development of a new product is a costly project [email communication]. The company’s standard cryocooler can handle 90\,W of excess thermal energy. Interestingly, a \textit{Pulse-pumped CW tunable Ti:sapphire laser} was reported at the CLEO conference (2009) by Hsiao-hua Liu from KMLabs~\cite{Liu:2009}.

\subsection{Advanced Photonic Sciences (APS)}
This US company\footnote{\url{www.apslasers.com}} produces cryogenic lasers but mainly Yb:YAG (not Ti:Sa) but their closed-cycle cooling system (not requiring liquid nitrogen refrigerant) may be of interest for the long term since it gives 1\,kW of cooling power at around 100\,K, which would otherwise imply a high usage of liquid nitrogen (LN) refrigerant. The company claims “world-record performance” for a cryogenic laser with a CW sustained average power of 963\,W. A system producing 300\,W is almost a standard product (Thor-300). The company personnel, with others, have written a 74-page open-access review in 2016~\cite{Mackenzie:2021}. This gives an historical perspective, from their point of view, and details of the physics underlying cryogenic lasers in general (but sapphire and Ti:Sa are incidental to the main theme and described as a ‘legacy laser material’!).

\subsection{FMB Oxford cryocooler}
%The most obvious way to cool the laser crystal is to mount it on the bottom of a LN reservoir in a cryostat, as in the early research papers [*]. 
The cooler from this company\footnote{\url{fmb-oxford.com}} provides 3\,kW of cooling power at 100\,K. The companies KMLabs and APS favor cryocoolers, presumably because they do not want to restrict the use to customers that have suitable refrigeration supplies. (The vibrations associated with mechanical cooling engines can be circumvented by switching them off temporarily to interleave periods of cooling and operation.) For MAGIS-100 and AION, however, the supply of refrigerant is probably an incidental expense and the infrastructure for LN storage and transport would not be a major issue at a large facility such as Fermilab. Thus, it may be worthwhile to investigate this FMB Oxford cryocooler --- this company originated as part of Oxford Instruments (Synchrotron Group, sold in 2001) and is located near Oxford. The system is designed to cool crystals in monochromators for X-rays from synchrotrons with special attention to low-vibration operation (suppressing mechanical resonances and low turbulence flow). Thus, it is potentially suitable for laser crystals, although the tolerable level of vibration when amplifying narrow-bandwidth radiation would need to be quantified. Mechanical vibrations of relatively large objects (a few cm in size) are typically at frequencies that fall within the range that can be suppressed using fast negative feedback (bandwidths of MHz with suitable electronics). However, this requires either optical components that can handle hundreds of Watts (large diameter beams and suitable crystals) or negative feedback on the light before amplification (accounting for time delays).

\subsection{Diode Pumped Optical Laser for Experiments (DiPOLE) laser project}
\label{app:dipole_project}
A kilowatt-average-power pulsed system Yb:YAG laser cooled by recirculating helium gas was designed and constructed at the Central Laser Facility of the Science and Technology Facilities Council (STFC) at the Rutherford Appleton Laboratory\footnote{\url{https://www.clf.stfc.ac.uk/Pages/D100.aspx}}. It produces pulse energy of 150\,J at 1\,Hz, or 75\,J at 10\,Hz which corresponds to 750\,W of average optical power output. The system uses a slab geometry and has a large beam size ($75\times 75$\,mm), as is common for such pulsed laser facilities, and it is noteworthy for being cooled by high-pressure helium (10\,bar at 150\,K) from a specially designed cryocooler.

The recirculating helium gas is cooled by LN, via a heat exchanger, and is commercially available from GRE Ltd, Devon, UK\footnote{\url{gre-ltd.co.uk}}. This design with the crystal(s) in flowing helium gas is unlikely to be directly applicable to a narrow bandwidth CW case, but is included here as an example of an alternative approach to extracting the heat from the crystal surface. People from the Physics Department in Oxford were involved in the early stages.

\clearpage
\section{Pump lasers for Ti:sapphire laser}
\label{app:pump_lasers}

\subsection{Existing diode-pumped solid-state (DPSS) green lasers}
Currently, the laser systems in MAGIS-100 and AION are pumped by a newly developed Equinox system from M Squared Lasers (which has reliability issues). The Verdi laser from Coherent Inc.\ has been used successfully for many years (in Oxford and elsewhere). Whatever configuration is chosen, either coherent combination of many sources or a few much higher power amplifiers, it seems likely that a narrow bandwidth master laser will be required with the power that goes beyond what is available from diode lasers. The operation of a diode laser with extremely high-powered amplifiers would be a fragile setup. Therefore, at least one Verdi laser, or otherwise, is required to provide 18\,W at 532 nm. (A 25\,W DPSS laser is available from Spectra-Physics: Millennia laser\footnote{\url{https://www.spectra-physics.com/en/f/millennia-ev-cw-dpss-laser}}.)

\subsection[Experimental demonstration of 130 W at 532 nm]{Experimental demonstration of 130\,W at 532\,nm}
Extremely efficient generation of CW radiation at a wavelength of 532\,nm has been demonstrated that produced 134\,W from a fundamental power of 149\,W by second-harmonic generation in an external optical resonator around a lithium triborate crystal~\cite{Meier:2010}. This was limited by the available fundamental power and could be pushed to even higher powers.

Nowadays, the company Azurlight Systems produces $>100$\,W fiber lasers and lower power green lasers, but has a strong interest in developing higher power systems in the green [private communication]. This company has a good reputation for stable laser systems for use in cold-atom experiments, whereas other fiber laser companies are more focused on industrial applications.

\subsection{Doubled fiber lasers}
Fiber lasers commercially available from the company IPG\footnote{\url{https://www.ipgphotonics.com/en}} can reach 100\,W of CW output and are marketed mainly for industrial applications such as cutting and welding; they are not commonly used for pumping Ti:Sa so their suitability would have to be tested. Interestingly, IPG produces a quasi-CW laser that outputs 1\,kW of green light. At first sight, this pulsed laser might seem unsuitable, but the frequency of the pulses can be up to 250\,MHz; a repetition time of 4\,ns is much less than the response time of a laser cavity which therefore smooths the output intensity. A demonstration of this \textit{Pulse-pumped CW tunable Ti:sapphire laser} was reported at the CLEO conference (2009) by Hsiao-hua Liu from KMLabs~\cite{Liu:2009}, as noted in Appendix~\ref{app:kmlabs}. In Ti:Sa the upper laser level has a lifetime of around $3.2\,\mu\text{s}$ which is relatively short for solid-state laser, making relaxation oscillations less likely. Broadly speaking, although the lifetime for spontaneous emission is critical for determining whether CW population inversion can be obtained, it is not so important under operating conditions where the simulated emission rate greatly exceeds that of the spontaneous process. An important parameter is the lifetime of the photon in the optical cavity (cavity lifetime $\tau_{\text{cav}}$). The round-trip time is given by the speed of light divided by the length of a round trip, e.g.\ $L/c = 2$\,ns for $L = 0.6$\,m. The lifetime for photon in the cavity equals this multiplied by the number of round trips which can be approximated as the finesse, e.g.\ for an amplitude reflectivity of $R = 0.85$, the finesse
\begin{equation}
    \mathcal{F}= \pi\sqrt{R}/(1-R) = 20\ ,
    \notag
\end{equation}
hence $\tau_{\text{cav}} = 40$\,ns.

This system with a storage time long compared to the period between pump laser pulses could be analyzed using the usual set of coupled rate equations for the photon number (or intensity) in the cavity and population inversion density used to describe Q-switching in laser physics texts~\cite[pp.~188--193]{Hooker:2010}, but that is beyond the scope of this report. The likely conclusion of such an analysis is that there is a benefit in making the cavity longer to increase the photon lifetime. A (folded) cavity length of tens of meters could be achieved, increasing the cavity lifetime to $2\,\mu\text{s}$ which smooths out fluctuations on shorter timescales. Even with such a long response time compared to the smoothing, however, this seems a risky approach for a system that must also has a narrow bandwidth, although the bandwidth requirement becomes less stringent in inverse proportion to the pulse length.

\clearpage
\section{‘Cost benefit’ analysis}
\label{app:cost_benefit}

This preliminary survey of various possibilities indicates possible choices such as the coherent addition of many tens of standard room temperature laser systems ($< 10$\,W each) or higher power ($>100$\,W)  systems (e.g.\ cryocooled lasers). Currently it appears that the former approach is the default in MAGIS-100 and AION and it remains to be seen how far it can be pushed in terms of reliability and cost (any of these approaches is likely to exceed 1 million currency units: \$, £ or €). Cryocooling is a ‘standard’ technique in kilowatt laser systems but requires a bespoke design and the associated research and development effort.

\end{appendices}

\end{document}